# Understanding Transient Photoluminescence in Halide Perovskite Layer Stacks and Solar Cells

*Lisa Krückemeier\*, Benedikt Krogmeier, Zhifa Liu, Uwe Rau and Thomas Kirchartz\**

**Affiliations**
L. Krückemeier, B. Krogmeier, Z. Liu, Prof. U. Rau, Prof. T. Kirchartz
IEK5-Photovoltaik
Forschungszentrum Jülich
52425 Jülich, Germany
E-mail: l.krueckemeier@fz-juelich.de
E-mail: t.kirchartz@fz-juelich.de
L. Krückemeier, Prof. U. Rau
Faculty of Electrical Engineering and Information Technology, RWTH Aachen University, Mies-van-der-Rohe-Straße 15, 52074 Aachen, Germany
Prof. T. Kirchartz
Faculty of Engineering and CENIDE
University of Duisburg-Essen
Carl-Benz-Str. 199, 47057 Duisburg, Germany

**Abstract**

While transient photoluminescence measurements are a very popular tool to monitor the charge-carrier dynamics in the field of halide perovskite photovoltaics, interpretation of data obtained on multilayer samples is highly challenging due to the superposition of various effects that modulate the charge-carrier concentration in the perovskite layer and thereby the measured PL. These effects include bulk and interfacial recombination, charge transfer to electron or hole transport layers and capacitive charging or discharging. Here, numerical simulations with Sentauraus TCAD, analytical solutions and experimental data with a dynamic range of ~7 orders of magnitude on a variety of different sample geometries from perovskite films on glass to full devices are combined to present an improved understanding of this method. A presentation of the decay time of the TPL decay that follows from taking the derivative of the photoluminescence at every time is proposed. Plotting this decay time as a function of the time-dependent quasi-Fermi level splitting enables distinguishing between the different contributions of radiative and non-radiative recombination as well as charge extraction and capacitive effects to the decay.

## 1. Introduction

Technological development of halide-perovskite solar cells towards even higher efficiencies requires ways of understanding and quantitatively analyzing the main loss processes.[1-3] Non-radiative recombination is one of the main loss processes in basically any solar-cell technology[4, 5] including perovskites that leads to reduced open-circuit voltages at a given illumination condition.[6-9] Transient photoluminescence (TPL) is a frequently used tool to monitor the charge-carrier dynamics and investigate these recombination losses.[10-12] While TPL measured on bare perovskite films on glass is a well-understood and frequently-used method to derive charge-carrier lifetimes and recombination coefficients,[13, 14] it is typically the recombination at interfaces between the absorber and charge transfer layers that is the dominant sources of recombination in a complete device.[15-21]

However, adding contact layers to the perovskite film not only adds additional recombination paths but also leads to effects like charge-carrier separation or interface-charging effects[21-23] that may change the recombination kinetics fundamentally. Thus, the information on recombination may be totally obscured by interfacial effects such that, e.g., a faster photoluminescence (PL) decay cannot anymore be interpreted simply as an increased recombination coefficient. Hence, misinterpretation of experimental data becomes likely, especially if the information contained in the decay curve is reduced to a single value – the characteristic decay time of a mono-exponential decay. The present paper introduces a method to analyse the differential PL decay that uses the derivative of the photoluminescence at every time during the transient.[22] We propose to plot this decay time as a function of the corresponding quasi-Fermi level splitting allowing us to better understand the complex interplay between charge extraction and interface or contact charging with radiative and non-radiative recombination. We use a combination of numerical simulations with Sentauraus TCAD, analytical models and experimental data to illustrate how the different effects influence the PL transients and the resulting decay times.

In the following, we investigate transient photoluminescence measurements on the different sample geometries shown in the overview in **Figure 1** that start with films on glass and continue via layer stacks to full devices and discuss their respective peculiarities. We show how the addition of further layers and interfaces modifies the transients and adds physical effects that must be considered. With this step-by-step description we aim to create an understanding of which processes dominate and are important for these different sample types and how they affect the transient PL decay and the extracted decay time. This step-by-step approach is conceptually similar to the already well-understood analysis of steady state PL (SSPL) as a function of sample type that allows screening contact layers for minimum recombination losses[15, 24] and to also analyze and quantify changes in ideality factor[20, 24] and resistive effects.[25] In SSPL any change in absolute intensity between different samples suggests that the difference between the samples (e.g. an additional interface and layer) must be the cause of the loss in PL which is linked to a loss in quasi-Femi level splitting. In contrast, a clear correlation between the decay of a transient PL signal and the amount of recombination is not necessarily present, because in a transient experiment currents can temporarily flow even at open circuit.

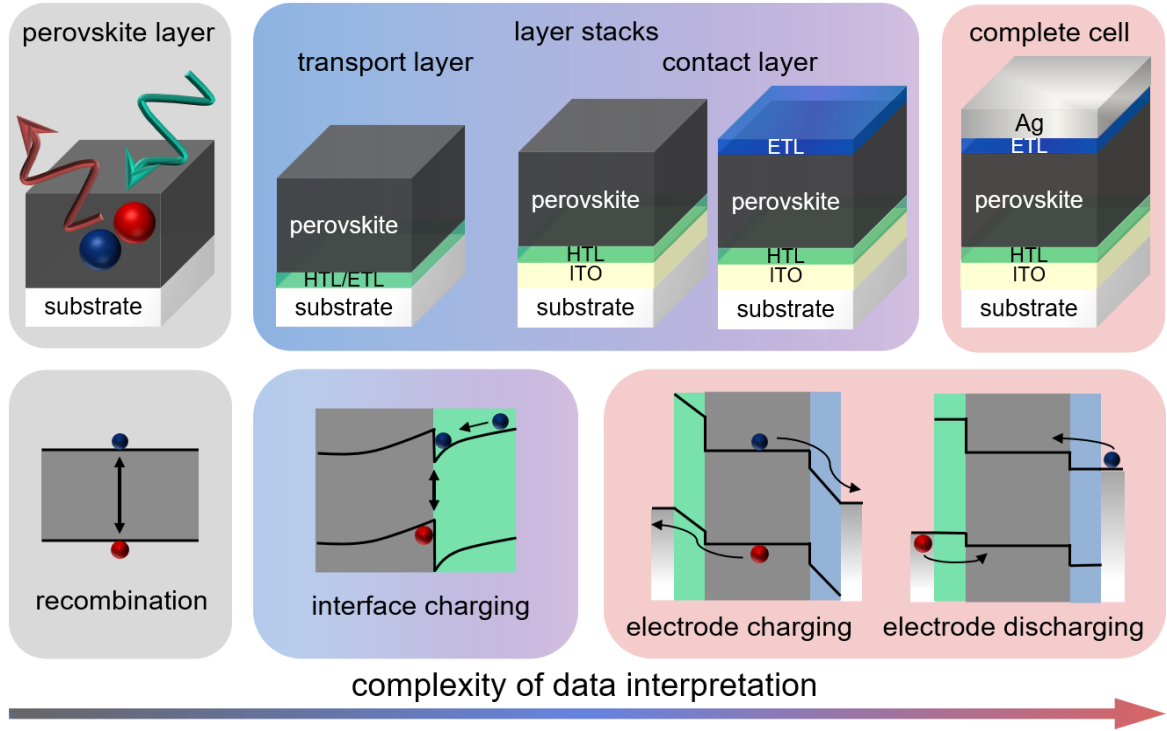

**Figure 1:** Overview over the different halide perovskite sample types that are investigated with transient photoluminescence. Depending on the sample type (i.e., absorber layer on glass or absorber layer with one or two contact layers attached), the complexity of the interpretation of data increases from left to right. The second row illustrates typical mechanisms that affect the PL decay in different sample geometries. While the film on glass is only affected by bulk and surface recombination, samples with interfaces to charge extraction layers are affected by charge accumulation and recombination at the interfaces between absorber and charge-extraction layer. On fully contacted devices, charging and discharging of the electrodes finally affects the global band diagram and therefore the transients measured on full devices.

## 2. Perovskite film on glass

The simplest sample geometry is that of a perovskite film on glass. Therefore, we want to use this device geometry as a starting point to recapitulate the basic theory of TPL measurements and data interpretation that is already quite well understood and briefly introduce the terminology and the physical parameters that we will use during the discussion. Furthermore, we introduce the representation of the decay time $\tau_{TPL}$, which results from the derivative of the photoluminescence with respect to time that we plot as a function of the corresponding time-dependent quasi-Fermi level splitting $\Delta E_F$.

## 2.1. Perovskite film on glass with a passivated surface

At first, we consider charge-carrier recombination only in the bulk and leave out the effect of surface recombination, which is discussed separately in section 2.2. A perovskite film on glass whose surface is passivated, e.g. by an organic passivation layer,[26, 27] is suitable as a corresponding test structure, because the interface between perovskite layers and glass substrates is typically quite inert and not particularly recombination active.[13, 28] When the semiconducting perovskite film is photoexcited by a laser pulse, free excess-electrons $\Delta n$ and equal density of holes $\Delta p = \Delta n$ are generated. The initial generation profile of these excess-charge carriers depends on the optical properties of the perovskite, its thickness, and the excitation wavelength. The initially inhomogeneous distribution of the charge-carrier concentration is flattened by diffusion. Since the perovskite film is typically only a few hundred nanometers thick and the charge-carrier mobility is typically > 1 cm²/Vs,[29] this carrier equilibration occurs in the first hundreds of pico- to several nanoseconds after the laser pulse. For simplicity, we therefore assume that the system has no spatial gradients of electron or hole concentrations or the electrostatic potential. This assumption leads us to the convenient situation that diffusion and drift currents can be neglected and that the photogenerated charge carriers $\Delta n$ vanish over time only through different recombination processes. The rate equation

$$\frac{\partial \Delta n}{\partial t} = -k_{\text{rad}}\left(np - n_i^2\right) - \frac{np - n_i^2}{\tau_p n + \tau_n p} - C_n n\left(np - n_i^2\right) - C_p p\left(np - n_i^2\right) \tag{1}$$

accounts for the rates of these different competing processes and describes the change of excess-charge carrier concentration as a function of time. Here, $n_i$ is the intrinsic charge-carrier density, $n$ is the total electron and $p$ the total hole concentrations, being defined as $n = n_0 + \Delta n$ and $p = p_0 + \Delta n$, respectively, with the corresponding equilibrium concentrations $n_0$ and $p_0$. For the moment, we keep the condition $\Delta p = \Delta n$ and therefore do not distinguish between the rate equations for electrons and holes. For a perovskite film on glass three different bulk-recombination mechanisms need to be considered, which are radiative band-to-band recombination[30, 31] and non-radiative recombination via Auger[32] or first-order Shockley-Read Hall (SRH).[33, 34] The radiative recombination coefficient is given by $k_{\text{rad}}$, $\tau_n$ and $\tau_p$ are the non-radiative SRH lifetime for electrons and holes, respectively. For the Auger recombination rates, $C_n$ represents the Auger coefficient for electrons and $C_p$ is the Auger coefficient for holes. Note that the effect of photon recycling[35] is not explicitly considered in this paper, so the radiative coefficient $k_{\text{rad}}$ is the corresponding external quantity. Photoluminescence is based on measuring the emitted photons that have been created by radiative recombination with a rate proportional to $np$. Accordingly, the PL intensity $\phi_{\text{TPL}}$ depends on the product $np$ of electron and hole concentrations. The time-resolved photoluminescence itself contains information about these entire recombination processes because the decrease in $n$ and $p$ results from the sum of all recombination processes.

While the radiative recombination coefficient and the Auger coefficients are important material properties, the SRH lifetimes can vary substantially from sample to sample within a material system. This is because SRH lifetime depend on defect concentrations which may vary e.g. with processing conditions. Hence, while the determination of all recombination coefficients is important for a material system, the SRH lifetimes are the key sample dependent property that might vary in a series of samples of the same material. Hence, TPL measurements on semiconducting materials would very often be done to extract the SRH lifetimes of individual samples. Given that the PL transient is affected by a several recombination mechanisms, it is necessary to briefly discuss how to extract the SRH lifetimes from the PL transients and to establish a common terminology that we use for the remainder of the article.

The term charge-carrier lifetime in general describes the dependence of the electron or hole concentration as a function of time after the generation of electrons and holes (e.g. by a laser pulse) has stopped. Often, the characteristic decay time constant $\tau_{mono}$ of an exponential fit of the form $\phi_{TPL} = \phi_{TPL}(0)\exp(-t/\tau_{mono})$ to TPL data is also called a 'lifetime'. This lifetime concept is meaningful and suitable for a doped semiconductor film, crystal or wafer that is operated in low-level injection (less injected minority carriers than pre-existing majority carriers). In this case all recombination mechanisms stated in Equation 1 become linear in minority carrier concentration and a lifetime (often specifically called effective lifetime $\tau_{eff}$) could easily be defined as the decay time of minority carrier decay. I.e. if electrons are minority carriers their decay would be mono-exponential and follow $\Delta n = \Delta n(0)\exp(-t/\tau_{eff})$ no matter what recombination mechanism dominates[36] and its effective lifetime $\tau_{eff}$ is a constant, carrier concentration independent value (Supporting Information Table II).

Lead-halide perovskite films, however, are typically intrinsic enough that during a TPL experiment both types of carriers are present in roughly equal concentrations. In this situation, called high-level injection (HLI), both types of carriers have a finite lifetime. Let us assume for simplicity that electron and hole concentrations are exactly equal, i.e. $n = p$, to illustrate the difference between the time-dependent decay time and a lifetime. In this case, the rate equation (Equation 1) can be simplified to

$$\frac{dn(t)}{dt} = -(C_n + C_p)n(t)^3 - k_{rad}n(t)^2 - \frac{n(t)}{(\tau_n + \tau_p)} = -C_{Auger}n(t)^3 - k_{rad}n(t)^2 - \frac{n(t)}{\tau_{SRH}^{bulk}}, \qquad (2)$$

which can be solved analytically. The respective result and solutions for further problems are listed in Table S1 in the Supporting Information. Here, we only want to have a look at the illustrated solutions being shown in **Figure 2**a. In high level injection, different recombination mechanisms will lead to differently shaped transient decays. Among the three typically studied bulk-recombination mechanisms (SRH, radiative and Auger) only SRH would lead to a mono-exponential decay. Hence, the situation where only SRH recombination occurs can be described by a single lifetime value that is given for $n = p$ by the sum $\tau_{SRH}^{bulk} = \tau_n + \tau_p$. The dotted lines in Figure 2a represent this case for three different SRH lifetimes, namely 100 ns (yellow), 500 ns (red) and 2 µs (blue). Both, Auger, and radiative recombination lead to faster initial decay, with a higher radiative recombination coefficient $k_{rad}$ and higher Auger coefficients $C_n$ and $C_p$ enhancing the effect. As an example, the $k_{rad}$ is varied in Figure 2b for the SRH lifetime of 2 µs. At longer times, the slope approaches the one from SRH- only scenario.

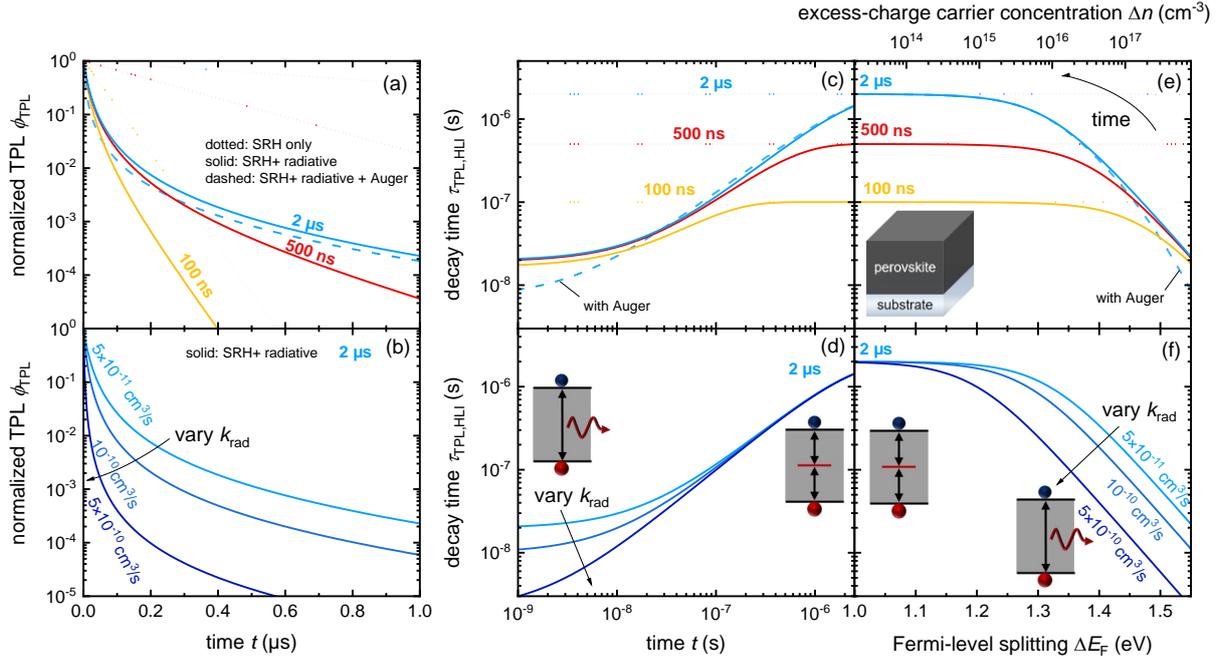

**Figure 2:** Illustration of the analytical solutions of the differential equations for high-level injection describing an undoped perovskite film on glass with passivated surface. (a-b) Normalized transient photoluminescence decays $\phi_{TPL}$, starting from similar initial charge-carrier concentration, for (a) varying SRH lifetimes with $\tau_{SRH}^{bulk} = \tau_n + \tau_p$ equals 100 ns, 500 ns or 2 µs, and (b) varying radiative coefficients $k_{rad} = 5\times10^{-11}\,cm^3s^{-1}$, $5\times10^{-10}\,cm^3s^{-1}$ or $10^{-10}\,cm^3s^{-1}$. Different scenarios are selected to differentiate the influence of the respective recombination mechanisms on the decay. (c-d) presents the decay time $\tau_{TPL,HLI}$ in HLI as a function of time $t$ or in (e-f) plotted versus quasi-Fermi-level splitting $\Delta E_F$ which is linked to the excess-charge carrier concentration via Equation 4 using $n_i = 8.05\times10^4\,cm^{-3}$.[13]

Consequently, if any other recombination mechanism plays a role, the decay is no longer mono-exponential and cannot be described by a single lifetime. Thus, we promote the use of a carrier-density-dependent decay time defined by

$$\tau_{TPL} = \left(-\frac{1}{m}\frac{d\ln(\phi_{TPL})}{dt}\right)^{-1} \qquad (3)$$

to interpret TPL decays, with the factor $m$ relating to the injection level ($m = 1$ for low-level injection (LLI), $m = 2$ for high-level injection (HLI)). Figure 2c-d display $\tau_{TPL,HLI}$, resulting from the analytical solution of the TPL decays in Figure 2a-b, versus time. Figure 2e-f presents these decay times as a function of the quasi-Fermi-level splitting $\Delta E_F$

$$\Delta E_F = kT\ln\left(\frac{\Delta n^2}{n_i^2}\right). \qquad (4)$$

Plotting the decay time $\tau_{TPL}$ versus $\Delta E_F$ allows more easily distinguishing the influences of the individual processes because they typically differ in their dependence on carrier density (the different recombination mechanisms) or external voltage (electrode discharging), while they do not have a fundamental time dependence. A similar method of representation – i.e. time constant vs. voltage – is also typically used for the analysis of transient photovoltage (TPV)[37-39] and open-circuit voltage decay (OCVD)[40-42] measurements. These measurements also allow deriving differential time decay that are however frequently denoted as "lifetimes" in the literature.[39, 41, 43, 44] Radiative and Auger recombination define the shape of the decay time at high Fermi-level splitting, i.e. directly after the laser pulse. The comparison of the analytical solution with and without Auger recombination illustrates that Auger recombination steepens the increase of $\tau_{TPL}$. In all cases, the saturation of the decay time on the value of the SRH lifetime is clearly visible (Figure 2a, c-d). The choice of $m = 2$ for $n = p$ ensures that the decay time given by Equation 3 will saturate towards $\tau_{TPL,HLI} = \tau_n + \tau_p$ for long times or low $\Delta E_F$.

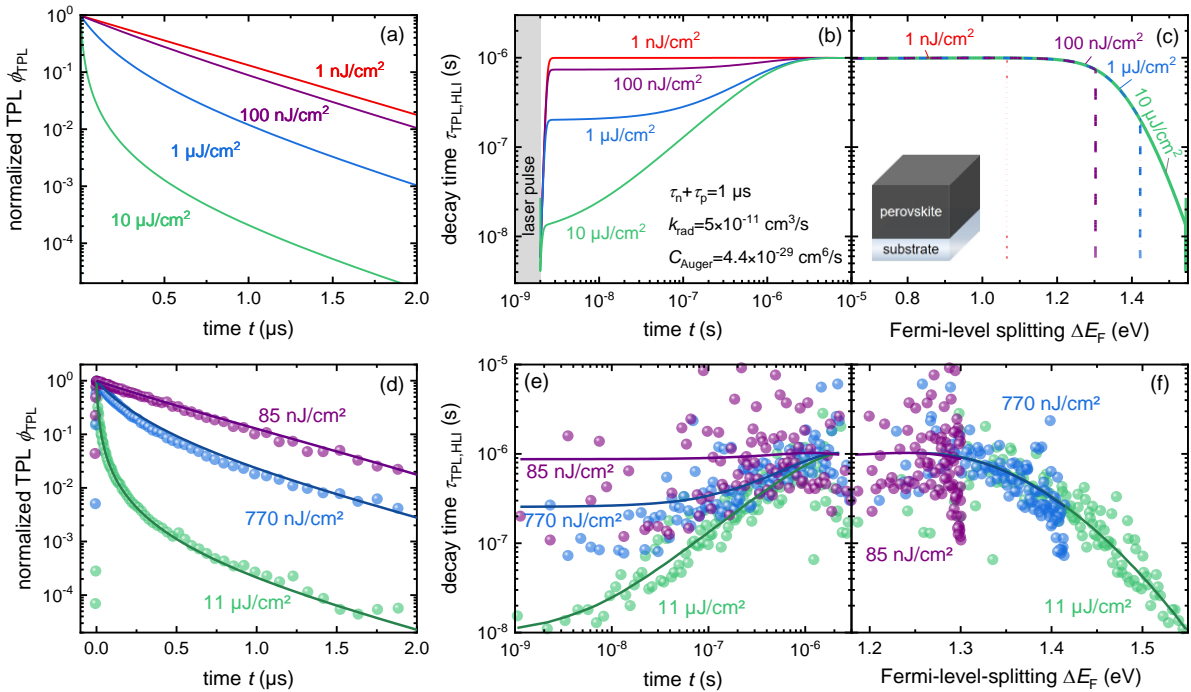

**Figure 3:** (a-c) Simulated data (Sentaurus TCAD) of transient photoluminescence measurements of a $CH_3NH_3PbI_3$ perovskite film with $\tau_n = \tau_p = 500$ ns ($\tau_{SRH,bulk} = \tau_n + \tau_p = 1$ μs), $k_{rad} = 5 \times 10^{-11}$ cm$^3$s$^{-1}$ and $C_n = C_p = 4.4 \times 10^{-29}$ cm$^6$s$^{-1}$ for different laser excitation fluences (1 nJ/cm$^2$, 100 nJ/cm$^2$, 1 μJ/cm$^2$, 10 μJ/cm$^2$). (a) Normalized photoluminescence decays over time Sentaurus TCAD simulations on a semi-logarithmic scale. For low laser fluences or later times the exponential decay caused by SRH recombination ($R_{SRH} \propto \Delta n$) dominates the decay. At higher laser fluences higher order recombination mechanisms like radiative or Auger recombination dominate the decay. (b) Carrier-density-dependent decay time $\tau_{TPL,HLI}$ assuming high-level injection ($m=2$) determined via Equation 3 versus time and (c) versus Fermi-Level splitting. (d-f) Experimental data reused from ref.[13] of transient photoluminescence measurements of a $CH_3NH_3PbI_3$ perovskite film on glass for different laser excitation fluences (85 nJ/cm$^2$, 770 nJ/cm$^2$, 11 μJ/cm$^2$). (d) shows the normalized photoluminescence decays over time from measurements (open symbols). Solid lines represent global fits including radiative and trap-assisted SRH and Auger recombination. (e) Decay time $\tau_{TPL,HLI}$ assuming HLI ($m=2$) versus time and (f) versus quasi Fermi-level splitting.

This new method of representation ($\tau_{TPL}$ vs. $\Delta E_F$) becomes even more important if TPL measurements at different excitation fluences are compared, because the photoexcited charge-carrier density critically influences which recombination type dominates the TPL decay. **Figure 3** presents a comparison of the simulated situation (a-c) and experimental data (d-f) of a perovskite film on glass but now for varying laser fluences. Again, the results are shown as normalized TPL decays and their decay times over time or Fermi-level-splitting. Figure 2b and e reveal that using the time as x-axis is not very informative, because $\tau_{TPL,HLI}$ at a specific delay time after the pulse can have very different values depending on the excitation energy. Only when one plots the $\tau_{TPL,HLI}$ versus Fermi level splitting, which depends on the charge-carrier concentration, the information merges and the curves complement each other. Finally, it should be noted that both the Sentaurus simulation and the analytical solution reflect the trend of the experimental data and are therefore well suited to describe a pure, passivated perovskite film on glass.

### 2.2. Layer with surface recombination

TPL measurements on unpassivated perovskite films on glass are often used to characterize not only the bulk but also the surface. If the bulk properties of the perovskite are already known, this sample type can be used to extract the corresponding surface properties from the TPL measurement. In general, the rate of surface recombination may be limited either by the transport of charge carriers to the surface or by the surface recombination velocities $S_n$ and $S_p$ of electrons and holes at the interface itself. Due to the relatively low thickness of halide perovskite thin films (typically < 1μm) and the typically high mobilities ($\mu > 1 cm^2/Vs$), the transport of electrons and holes to the surface can be considered fast compared to the rate of recombination at the surface,[13] which can be quite low relative to many other semiconductors. Therefore, we do not have to distinguish between the average carrier concentration in the bulk of the perovskite-thin film and the carrier concentrations at either of the surface. This finding implies that the rates of bulk and surface recombination can just be added up and it is not necessary to calculate the concentration of electrons and holes at the film surface. The surface recombination rate per unit area is given by

$$R_{surf} = \frac{np - n_i^2}{n/S_p + p/S_n} \approx \frac{n}{1/S_p + 1/S_n} \tag{5}$$

whereby the approximation sign is valid for $n = p$ and $np \gg n_i^2$. Equation 5 is essentially a modified SRH recombination rate, where all instances of lifetime are replaced by inverse surface recombination velocities $S_n$ and $S_p$ of electrons and holes. This implies that $R_s$ is a rate per unit area and time rather than per unit volume and time as the SRH recombination rate in the bulk. If we additionally assume $S_n = S_p = S$ for simplicity, we have $R_{surf} = Sn/2$. If we want to combine the surface recombination rate per unit area with the bulk recombination rates per volume and time, we have to either multiply all volume recombination rates with the thickness $d_{pero}$ of the film or divide the surface recombination rate $R_{surf}$ by the thickness. Doing the latter, we could write the rate equation including surface recombination at one surface (the other being perfectly passivated) as

$$\frac{dn(t)}{dt} = -k_{rad}n(t)^2 - \frac{n(t)}{\tau_{SRH}^{bulk}} - \frac{Sn(t)}{2d_{pero}}. \tag{6}$$

We note that in Equation 6, the bulk SRH term and the surface recombination term are both linear in electron concentration which suggests that we could use the concept of surface lifetime $\tau_{surf}$ and an effective SRH lifetime

$$\tau_{SRH}^{eff} = \left(\frac{1}{\tau_{SRH}^{bulk}} + \frac{1}{\tau_{SRH}^{surf}}\right)^{-1} = \left(\frac{1}{\tau_{SRH}^{bulk}} + \frac{S}{2d_{pero}}\right)^{-1}, \tag{7}$$

which would allow us to rewrite Equation 7 to obtain

$$\frac{dn(t)}{dt} = -k_{rad}n(t)^2 - \frac{n(t)}{\tau_{SRH}^{bulk}} - \frac{n(t)}{\tau_{SRH}^{surf}} = -k_{rad}n(t)^2 - \frac{n(t)}{\tau_{SRH}^{eff}} \tag{8}$$

Equation 8 has an analytical solution being listed in Table S1 in the Supporting Information.

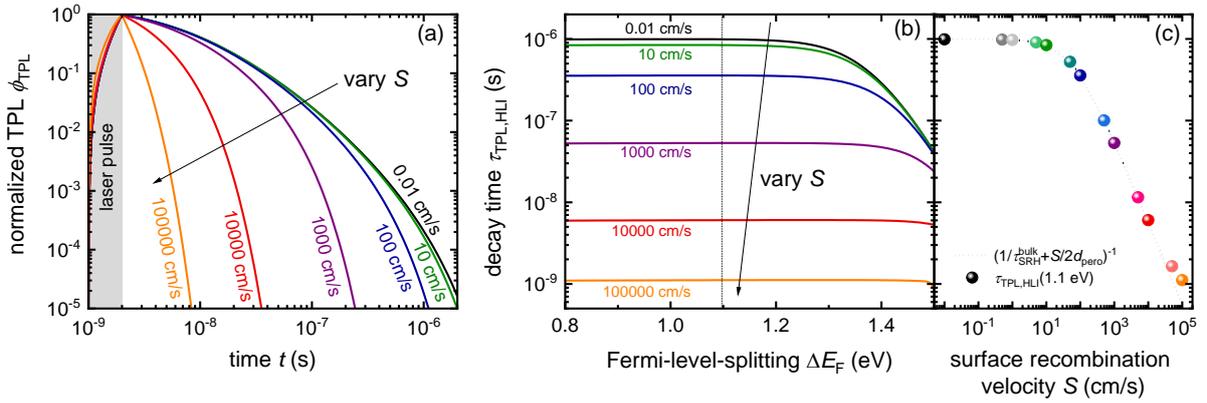

**Figure 4:** (a) Simulated normalized photoluminescence decays $\phi_{TPL}$ versus time of a $CH_3NH_3PbI_3$ perovskite film with a constant bulk SRH lifetime of $\tau_n = \tau_p = 500$ ns ($\tau_n + \tau_p = 1$ μs) and a varying surface recombination velocity $S$ (Sentaurus TCAD). A laser excitation fluences of 10 μJ/cm$^2$ was used. Further simulation parameters are listed in Table S3. The higher the surface recombination velocities the faster is the TPL decay. (b) Decay time $\tau_{TPL,HLI}$ over Fermi-Level splitting $\Delta E_F$ calculated form the normalized TPL decays via Equation 3. Figure 4c represents the correlation of the effective decay time on the surface recombination velocity $S$. The dotted line results from the analytical solution. The colored data points belonged to the saturated plateau values of $\tau_{TPL,HLI}$ at $\Delta E_F = 1.1$ eV from the numerical simulation.

To check whether the analytical description in Equation 7 is adequate, we have also simulated the corresponding structure of a perovskite film with one recombination-active surface numerically. The resulting PL transients and their decay times $\tau_{TPL,HLI}$ as a function of the quasi-Fermi-level splitting for various surface recombination velocities $S$ are summarized in **Figure 4**a and b. Faster surface recombination velocities $S$ lead to faster TPL decays and the saturation value of the decay time at low

Fermi-level-splitting decrease. These saturation values of $\tau_{\text{TPL,HLI}}$ match with the respective effective SRH lifetime $\tau_{\text{SRH,eff}}$. How well numerical and analytical solutions match is confirmed by Figure 4c, in which the corresponding plateau values of $\tau_{\text{TPL,HLI}}$ are compared with the calculated values using Equation 7. We conclude that the numerical results from Sentaurus TCAD and the analytical description fit well and only deviate for particularly large *S*. Therefore, the analytical model is suitable to analyse experimental TPL data of a perovskite film on glass with an unpassivated surface. The surface recombination velocity *S* can then be extracted from the plateau value of the decay time if the SRH bulk lifetime is already known from a measurement of a passivated sample.

### 3. Perovskite layer with charge extraction layer

Surface recombination is an important loss mechanism but usually the surface to ambient air is not of special interest. We are rather interested in the analysis of interfaces between the perovskite film and charge-extracting layers to evaluate and compare the quality of contact materials and assign recombination losses in the actual solar cell. There are a range of additional effects - not related to recombination - that will affect the shape of the TPL decay and that we will illustrate in the following using the example of a perovskite/PCBM bilayer. **Figure 5** a - d show band diagrams of a simulated TPL experiment on such a bilayer while Figures 5e - g show the TPL data at different fluences and the differential lifetime vs. time and vs. quasi-Fermi level splitting.

After the laser pulse has hit the sample, the equilibrium Fermi level $E_F$ splits up into quasi-Fermi-levels $E_{Fp}$ for holes and $E_{Fn}$ for electrons (see Figure 5b) and charge is transferred to the PCBM, which is visible by the increase of the Fermi-level split inside the PCBM. Due to the reasonably low mobility of the PCBM ($\mu_{\text{PCBM}} = 5\times10^{-2}$ cm²/Vs),[45] this increase is slower than in the perovskite and position dependent (compare Figure 5b with 5c). The injection of electrons into the PCBM leads to a reduction of the *np*-product in the perovskite and thereby to a reduction in the PL at early times that is not caused by recombination. In Figure 5c) (35 ns after the pulse), the quasi-Fermi levels are flat throughout the sample, while the bands are bend close to the interface. This bending implies that high densities of electrons accumulate in the PCBM close to the interface. This accumulation requires high fluences and low interfacial recombination velocities to happen and it implies that quantitatively describing interfacial recombination is only possible by including the effect of Poisson's equation in addition to the continuity equations for electrons and holes.[22] After the quasi-Fermi levels have equilibrated, recombination leads to a further reduction in PL. Depending on the properties of the bilayer, recombination may preferentially proceed (i) via interfacial recombination of electrons in the PCBM with holes in the perovskite, (ii) via re-injection of electrons from the PCBM into the perovskite and subsequent recombination in the perovskite. The first scenario is more likely if the interfacial recombination velocities are high and the conduction band offsets are large and the second scenario in the opposite case (low recombination velocities and small offsets). In both cases, additional aspects like the thickness $d_{PCBM}$ of the PCBM layer will affect the TPL decay.

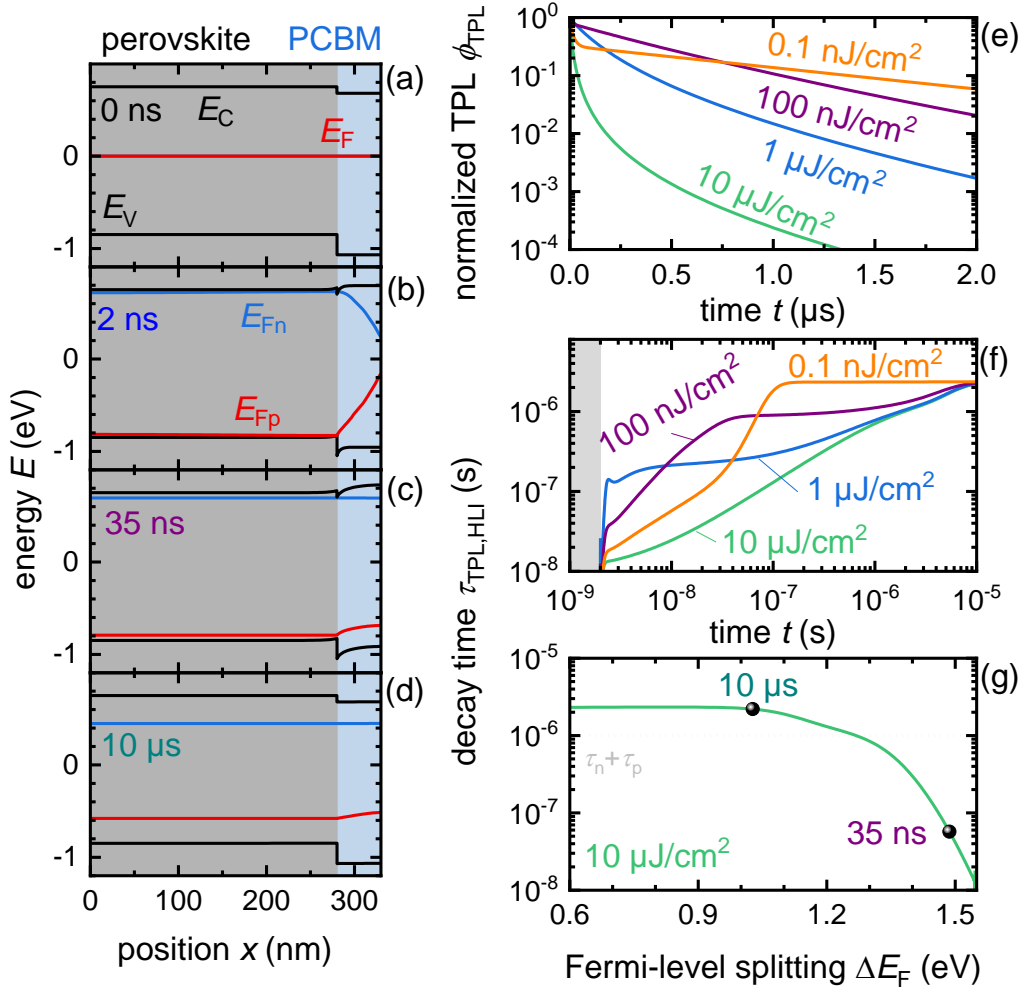

**Figure 5:** (a-d) Band diagrams of a perovskite film on glass with a charge-extracting PCBM layer on top before and at different time delays after the laser pulse excitation simulated with Sentaurus TCAD. (a) Equilibrium band diagram before the sample gets photoexcited. (b) shows the situations directly after the end of the laser pulse, when the quasi Fermi-level splitting in the perovskite is the highest but a negligible density of electrons has been transferred to the PCBM. In (c) substantial transfer of electrons to the PCBM has happened resulting in electron accumulation and band bending in the PCBM close to the perovskite/PCBM interface (35 ns). (d) after 10 µs this band bending has vanished again, and the Fermi-level splitting has visibly decreased. (e) normalized PL and (f) decay time over time for different laser fluences. In (g) the decay time $\tau_{TPL,HLI}$ for the highest fluence is shown again but here as a function of the average quasi-Fermi-level-splitting in the perovskite layer. The respective time steps for which the band diagrams are shown are highlighted. The SRH bulk lifetimes were set to $\tau_n + \tau_p = 1\,\mu s$ with $\tau_n = \tau_p$. The perovskite/PCBM junction is a type II heterojunction, the difference in electron affinity is 70 meV and the interface recombination velocity is set to $S = 0.01\,\mathrm{cm s^{-1}}$. Additional simulation parameters and are listed in Table S3 and a detailed description of the simulation implementation can be found in the Supporting Information.

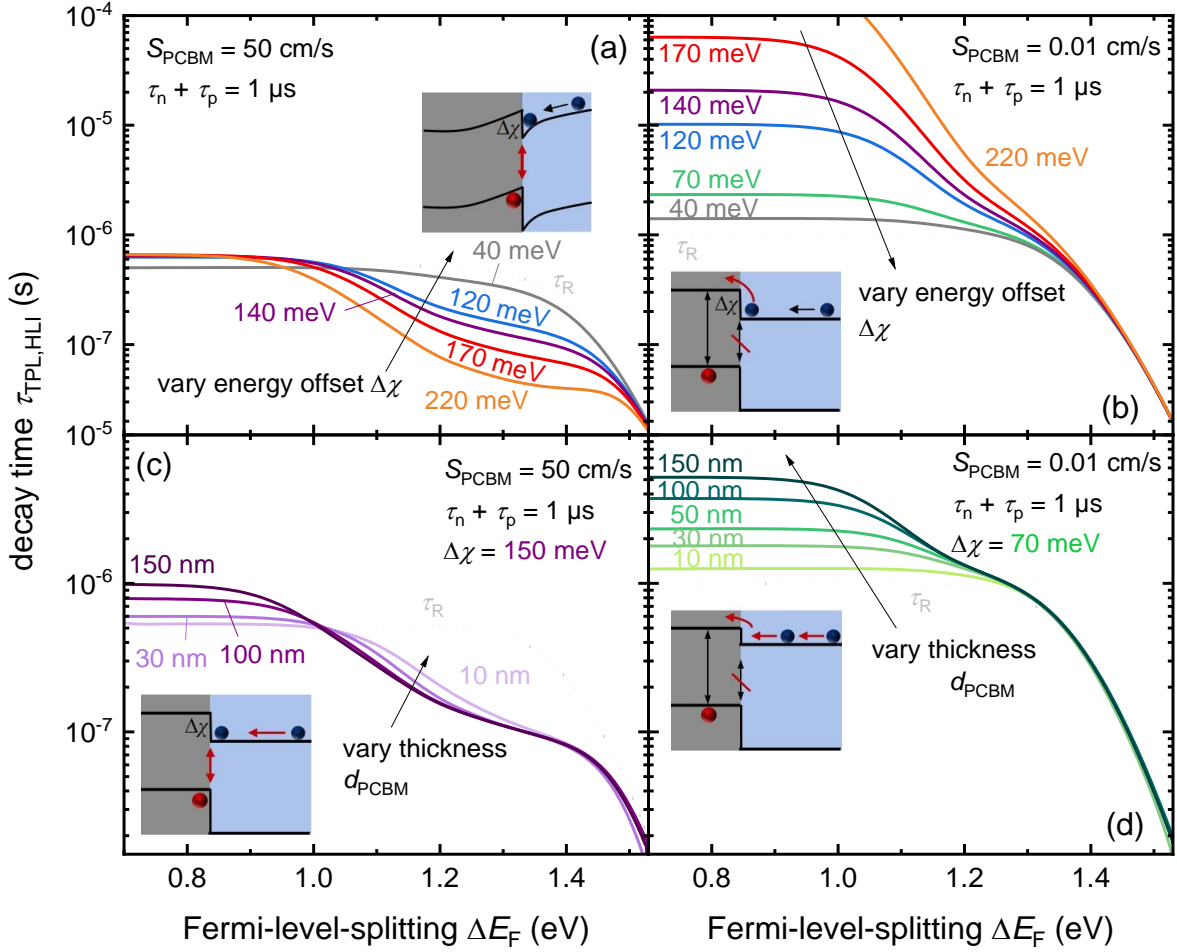

**Figure 6:** Decay time $\tau_{\text{diff,HLI}}$ versus Fermi-level splitting $\Delta E_F$ for the two scenarios, of (a, c) a high interfacial recombination velocity of $S_{\text{PCBM}} = 50$ cm/s at the perovskite/PCBM interface and (b, d) a negligible interfacial recombination velocity of $S_{\text{PCBM}} = 0.01$ cm/s. (a-b) shows a variation of the energy offset $\Delta\chi$ of the conduction bands and its impact on the decay time for each scenario and (c-d) a variation of the PCBM layer thickness $d_{\text{PCBM}}$. The dotted, grey lines show the simulation result of the pure perovskite film where only the combination of the different recombination processes define the course of the decay time, named recombination lifetime $\tau_R$ (details see SI). A comparison allows us to distinguish between pure recombination processes and the impact of additional effects e.g. interfacial charging, reinjection, or diffusion limitation. Additional simulation parameter can be found in Table S3 in the Supporting Information.

Based on **Figure 6**a and b we explain the impact of the conduction band offset $\Delta\chi$ (40 meV-220 meV), which is defined by the difference in electron affinity $\chi_{\text{PCBM}}$ of the PCBM and the perovskite $\chi_{\text{pero}}$, for the two scenarios. In this example the thickness $d_{\text{PCBM}}$ of the PCBM is set to 50 nm like in Figure 5. In Figure 6c-d the impact of a thickness change of the PCBM layer is illustrated. First, we discuss scenario (i) by showing decay times calculated for a non-negligible interface recombination velocity of $S = 50$ cm/s. We observe for all the different band offsets $\Delta\chi$ differently pronounced S-shaped curves. After the fast decay times at high $\Delta E_F$ due to charge transfer to the PCBM as well as the normal radiative and Auger recombination in the bulk, we observe an initial plateau of $\tau_{\text{TPL,HLI}}$ between around 1.2 and 1.4 eV quasi-Fermi level splitting. For even lower $\Delta E_F$, $\tau_{\text{TPL,HLI}}$ saturates at a second plateau. We observe that the first plateau nearly disappears for small offsets and is particularly pronounced in the opposite case. At an offset of 40 meV (grey line) the shape of the decay time approaches the simple case of an unpassivated perovskite film, when no additional effects besides recombination define the decay time (recombination lifetime $\tau_R$). The explanation for the first plateau is the effect of charge accumulation in the PCBM close to the interface. This effect reduces $\tau_{\text{TPL,HLI}}$ relative to longer times, where charge accumulation has been reduced to a point where all bands and Fermi levels are flat (see Figure 5d). The effect of charge accumulation gets more severe for higher band offsets, because the misalignment of the conduction bands leads to larger electron densities in the PCBM and reduces the rate of re-injection into the perovskite.

In addition, the thickness of the PCBM layer is varied between 10 – 150 nm for a medium-large energy offset of 150 meV. The respective decay times $\tau_{\text{TPL,HLI}}$, being presented in Figure 6c in different shades of purple with lighter colors representing thinner layers, point out that the first plateau gets more pronounced for thicker PCBM layers. Higher PCBM thickness also allows more impact of band bending, because of electrostatic arguments. For a given charge-carrier density, any change in electrostatic potential of $k_B T/q$ will occur over one Debye length, where $k_B$ is the Boltzmann constant, $T$ the temperature and $q$ the elementary charge. Hence, thicker layers will allow a larger change in electrostatic potential over the PCBM and therefore stronger band bending. The saturation value of $\tau_{\text{TPL,HLI}}$ (second plateau) at long times and small $\Delta E_F$ is also strongly affected by PCBM thickness. At longer times, recombination will be limited by the time the electrons take to diffuse through part of the PCBM layer before they can recombine at the interface. This effect leads to longer $\tau_{\text{TPL,HLI}}$ at small $\Delta E_F$.

Figure 6b shows the case, where interfacial recombination is negligibly small and the decay at longer times must be due to re-injection of electrons into the perovskite absorber layer (scenario (ii)). Now, $\tau_{\text{TPL,HLI}}$ at longer times or small $\Delta E_F$ can substantially exceed the bulk SRH lifetime of 1 μs. This is because electrons are injected into the PCBM and cannot escape from there other than by re-injection over a barrier of a certain height. This re-injection process is the rate limiting step at longer times and therefore leads to prolonged decay times at small $\Delta E_F$. Again, PCBM thickness has the same effect as before for the saturation value of the decay time. Larger thicknesses lead to longer $\tau_{\text{TPL,HLI}}$ due to the diffusion time through the PCBM (Figure 6d).

The effect of band bending, and thus charge-carrier dynamics in a perovskite layer in contact with a charge extraction layer cannot easily be described by an analytical solution anymore. Just for thin charge-extracting layers with small energy offset, it is still reasonable to use the saturation value of the decay time $\tau_{\text{TPL,HLI}}$ at small $\Delta E_F$ as effective lifetime and use Equation 7 (section 2.2) for estimating interface recombination velocities and bulk SRH lifetimes. Only then does the decay time roughly correspond to the pure recombination lifetime $\tau_R$. When interpreting data from TPL measurements on samples with thick transport layers or those with high band offsets, the presented effects, and their impact on the decay time $\tau_{TPL,HLI}$ cannot be ignored. Otherwise one assigns interface and the bulk lifetimes, which can differ by orders of magnitude from the actual one.

## 4. Perovskite layer with charge extraction layer and electrode

In the previous section, we have discussed the relevant case of a type II heterojunction between the perovskite and a charge-transport layer – in this case the ETL. Similar simulations could of course be done for hole transport layers that form a type II interface with the perovskite. Two aspects are however missing in section 3 that may appear in practice. One is the case where type I heterojunction is formed, and the second aspect is that an electrode (anode or cathode) may be present in addition to the ETL or HTL. In this section we are combining these two issues and simulate the layer stack glass/ITO/PTAA/perovskite. UPS measurements suggest that PTAA may have a negligible valence band offset to MAPI at least for the samples prepared with the recipe presented in ref.[46]. As in the previous section, we first have a look at the corresponding band diagram in equilibrium and at different delay times after laser pulse excitation being represented in **Figure 7**a-d. When comparing the equilibrium-band diagrams of Figure 5a and 7a, it is directly apparent that the equilibrium Fermi-level $E_F$ is no longer roughly in the middle of the perovskite band gap but is rather pinned by the ITO contact. When different layers are brought into contact charge-carrier injection from layers with higher electron (or hole concentration) into layers with lower concentrations occurs. Because ITO is much more conductive than PTAA and perovskite, the work function of the ITO sets the position of the Fermi level in the layer stack and holes are transferred into the PTAA and the perovskite. Therefore, the Fermi-level in the perovskite is much closer to the valence band edge $E_V$ than to the conduction band edge $E_C$, which implies that the equilibrium hole concentration $p_0$ is substantially higher than $n_i$. Thus, the situation is comparable to the one in a doped semiconductor even though the perovskite itself is assumed to be perfectly intrinsic.

This behavior caused by the ITO contact leads to a transition between high- and low- level injection during the TPL measurement. Directly after (Figure 7b) up to 3.93 µs after the laser pulse excitation (Figure 7c) roughly equal densities of electrons and holes are present in the perovskite ($n = p$, HLI), whereas for later times the density of electrons becomes much smaller compared to the one of holes ($p \gg n$, LLI) (see Figure 7d and Supporting Figure S10). Then the position $E_{Fp}$ relative to the valence band $E_V$ is approximately constant and the quasi-Fermi level splitting is reduced by a $E_{Fn}$ moving towards $E_{Fp}$. This change of injection level also has an impact on the recombination rates. Since the ratio of electrons and holes is not constant and also not explicitly known, it is no longer possible to simplify the differential equation in Equation 1 for high- or low-level injection (e.g.: $n = p$ or $p \gg n$).

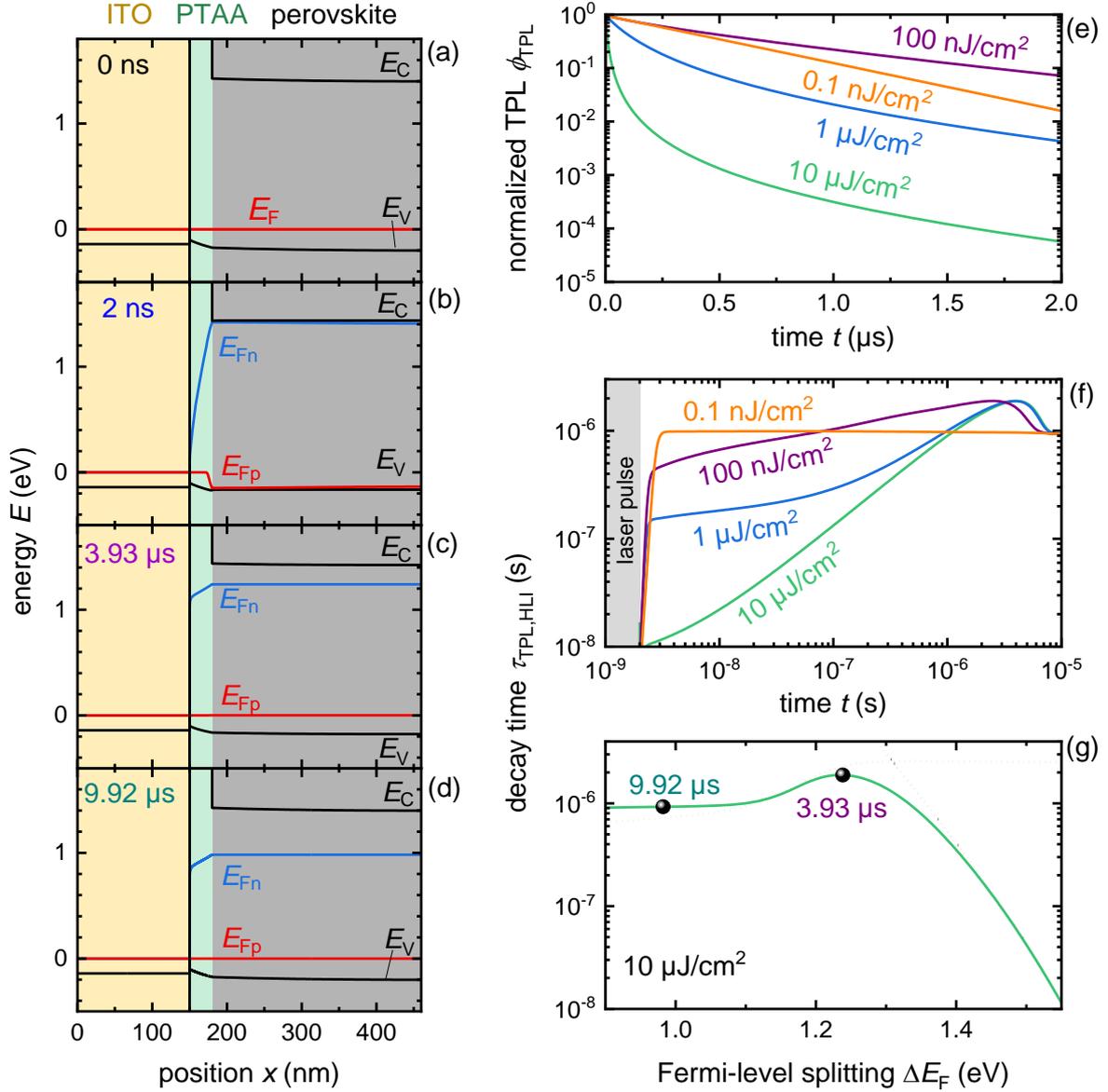

**Figure 7:** (a-d) Simulated band diagrams of a glass/ITO/PTAA/perovskite layer stack at different time delays before and after the laser pulses. These times are (a) before the laser pulse, (b) directly after the end of the laser pulse, (c) at a delay of 3.93 µs when substantial transfer of holes to the PTAA has happened, the quasi-Fermi-level for holes $E_{Fp}$ is flat and the perovskite is still in high-level injection and finally (d) after 9.92 µs. Then, the position $E_{Fp}$ relative to the valence band $E_V$ is approximately constant and the Fermi level splitting is reduced by a change of $E_{Fn}$. The concentration of electrons in the perovskite is much lower than the one for holes, corresponding to low-level injection. (e) Normalized PL transients and (f) decay time over time for different laser fluences. In (g) the decay time $\tau_{\text{diff,HLI}}$ for the highest fluence is shown again but here as a function of the average quasi-Fermi-level-splitting in the perovskite layer. The respective time steps for which the band diagrams are shown are highlighted. The perovskite/PTAA junction is a type I heterojunction and the interface recombination velocity are set to $S_n = S_p = 0.01 \text{ cms}^{-1}$. The SRH bulk lifetimes were set asymmetric to $\tau_n = 500$ ns and $\tau_p = 2$ µs enabling us to see the transition for HLI to LLI. Additional simulation parameters are listed in Table S3 in the Supporting Information.

This lack of information complicates the interpretation of the decay time in terms of bulk or surface lifetimes. In order to better see the transition from HLI to LLI, we use asymmetric SRH lifetimes in Figure 7 which are set to $\tau_n = 500$ ns and $\tau_p = 2$ µs. In high-level injection ($n \approx p$) the lifetime from SRH recombination is given by $\tau_n + \tau_p$=2.5 µs. At lower Fermi-level splitting (LLI), the situation in this example can be described in a simplified manner by the rate equation

$$\frac{dn(t)}{dt} = -\frac{n(t)}{\tau_n} \qquad (9)$$

with the solution $\Delta n = \Delta n(0)\exp(-t/\tau_n)$. With $p$ being approximately constant and fixed by the ITO work function, the photoluminescence flux will be proportional only to the minority carrier density, i.e. $\phi_{TPL} \propto n$. Thus, the decay time given by Equation 3 will give $2 \cdot \tau_n = 1$ µs, with the factor 2 originating from the choice $m = 2$ in Equation 3 that we use throughout the paper. We observe this transition from $\sim \tau_n + \tau_p$ to $\sim 2\tau_n$ when looking at the decay time curve vs. Fermi-level splitting (Figure 7g). A deviation from the value $\tau_{TPL,HLI} = \tau_n + \tau_p$ occurs because the transition from higher order recombination (radiative and Auger) to SRH recombination is not yet complete when the transition to LLI begins (indicated by the gray lines). In **Figure 8a** these findings are summarized by comparing the decay times for different ratios of electron and hole bulk lifetimes to highlight the transition from HLI to LLI. The dotted lines are a guide to the eye and mark the respective lifetime values of $\sim \tau_n + \tau_p$ (red) and $2\tau_n$ (blue). The dotted white lines show the simulations scenarios where only SRH or only radiative recombination is active. Note that these transitions from high- to low level injection that may occur due to doping of the absorber or Fermi-level pinning due to the (low or high) work function of a contact layer, create a dilemma for the definition of the decay time. We can either chose $m = 1$ or $m = 2$. No matter what we chose, it will affect the translation between decay time and lifetime in either of the two regimes.

Figure 8b shows the corresponding situation for the case, where interface recombination at the perovskite/PTAA becomes relevant. Here, we fix the bulk SRH lifetime at $\tau_n$ =500 ns and $\tau_p = 1$ µs as well as the interface recombination velocity for electrons at $S_n = 10^3$ cm/s and vary the interface recombination velocity $S_p$ for holes. Here, we also see a change of the decay time $\tau_{TPL,HLI}$ at a Fermi-level splitting of $\Delta E_F$ of around 1.2 eV due to the transition from high- to low-level injection. In HLI both interface recombination velocities $S_n$ and $S_p$ are relevant for the recombination rate, whereas in LLI only $S_n$ plays are role. Thus, at very low Fermi-level energies, the three curves converge to the same decay time value. Since the bulk lifetimes are high the interface recombination is the dominant non-radiative process and $\tau_{diff,HLI}$ should saturate to $\sim 2d_{pero}/S_n$=56 ns but the actual value is almost twice as high. In addition, we would not expect to see any transition for the case of symmetrically chosen lifetimes $\tau_n = \tau_p$ and surface recombination velocities $S_n = S_p$. The reason why this decay time at very low Fermi-level splitting differs, is the influence of space charge. In low-level injection, the high work function of the ITO leads to the creation of positive charge due to free holes inside the PTAA and the perovskite that is counterbalanced by negative charge at the ITO surface. This positive charge in the perovskite leads to band bending and therefore a slightly higher concentration of holes and thereby lower concentration of electrons at the PTAA/perovskite interface as compared to the bulk of the perovskite. At the PTAA/perovskite interface, electrons are the minority carriers in low-level injection and hence the reduction of electron concentration leads to a slight increase in the decay time in low-level injection relative to high-level injection (see Supporting Figure S10).

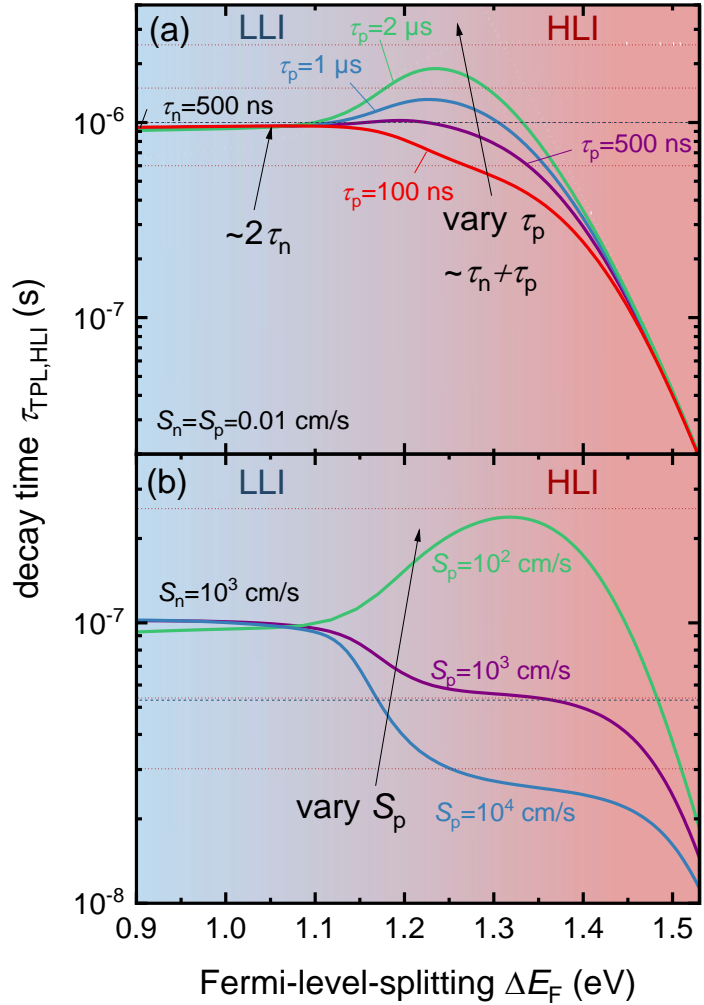

**Figure 8:** Decay times vs. Fermi-level splitting resulting from drift-diffusion simulation with Sentaurus TCAD of transient PL of glass/ITO/PTAA/perovskite stack. (a) Variation of the SRH bulk lifetime for holes $\tau_p$ from 100 ns (red) to 2 µs (green) at a constant electron lifetime $\tau_n$=500 ns, while the interface recombination velocity is negligibly slow ($S_n = S_p$= 0.01 cm/s). (b) Variation of the interface recombination velocity $S_p$ for holes at a constant bulk lifetime and recombination velocity for electrons of $S_n$=1000 cm/s. Both changes translate into observable changes in the decay time and illustrate the transition for high- to low level injection. Additional simulation parameters are listed in Table S3 in the Supporting Information.

## 5. Complete device stack

In case of the TPL of complete perovskite solar cells, the charging and discharging of the device electrodes adds an additional effect modifying the shape of the TPL decay. The mathematical problem can still be approximately described by an ordinary differential equation in time without any spatial dependences. This differential equation (neglecting Auger recombination for simplicity) could be written as

$$\frac{dn(t)}{dt} = -k_{rad} n(t)^2 - \frac{n(t)}{\tau_{SRH}^{eff}} - \frac{C_{area}}{qd_{pero}} \frac{dV(t)}{dt} \quad , \tag{10}$$

where $C_{area}$ is the area-related capacitance in F/cm² and $V$ the voltage between the electrodes. If we assume that the electron and hole densities scale with voltage as $n(t) = p(t) = n_i \exp(qV(t)/2k_B T)$, we can solve Equation 10 analytically. The effect of the last term in Equation 10 is to take into account that electrons will diffuse from the perovskite through the ETL to the cathode (Ag) and holes through the HTL to the anode (ITO) and change the surface-charge density on cathode and anode until the external voltage $V_{ext} = \Delta E_{F,ext}/q$ and the internal quasi-Fermi splitting $\Delta E_{F,int} = qV_{int}$ have equilibrated and no further current is flowing. At longer times, the electrons and holes may flow back from the electrodes through the ETL and HTL to the perovskite leading to long $\tau_{TPL,HLI}$ limited by the electrode capacitance and the differential resistance of the solar cell. The decay time $\tau_{TPL,HLI}$ at low $\Delta E_F$ is an $RC$ time constant with the $C$ being formed by the electrode capacitance and the $R$ being determined by the recombination resistance of the solar cell that increases exponentially towards smaller voltages as predicted by the diode equation. Hence, $\tau_{TPL,HLI}$ is also increasing exponentially towards smaller $\Delta E_F$ as previously observed for large and small signal $V_{oc}$ decays.[40, 41, 43, 44]

To illustrate the effects occurring in complete devices, **Figure 9**a - f presents the band diagrams simulated, using the full differential equations which include charge transport, at different delay times during a TPL measurement. Figure 9b depicts the band diagram directly after the pulse has hit the sample showing a substantial internal quasi-Fermi level splitting in the perovskite layer while the external voltage is still zero (see the Fermi levels in the ITO relative to the one in the Ag in Figure 9b). Figure 9c pictures the situation 5 ns after the pulse. At this stage, the quasi-Fermi levels are substantially split in the perovskite and also an external voltage has now been built up. However, there is still a substantial difference between the quasi-Fermi level splitting inside the perovskite absorber and the external quasi-Fermi splitting $\Delta E_{F,ext} = qV_{ext}$. Furthermore, there are large gradients $dE_{Fn}/dx$ and $dE_{Fp}/dx$ in the quasi-Fermi levels inside the electron and hole transport materials. These gradients are necessary to drive the electron- $J_n = n\mu_n\, dE_{Fn}/dx$ and hole-current density $J_p = n\mu_p\, dE_{Fp}/dx$ that have to flow from the absorber to the contacts to change the amount of charge stored on the electrodes. Hence, the average internal voltage is still substantially higher than the external voltage.

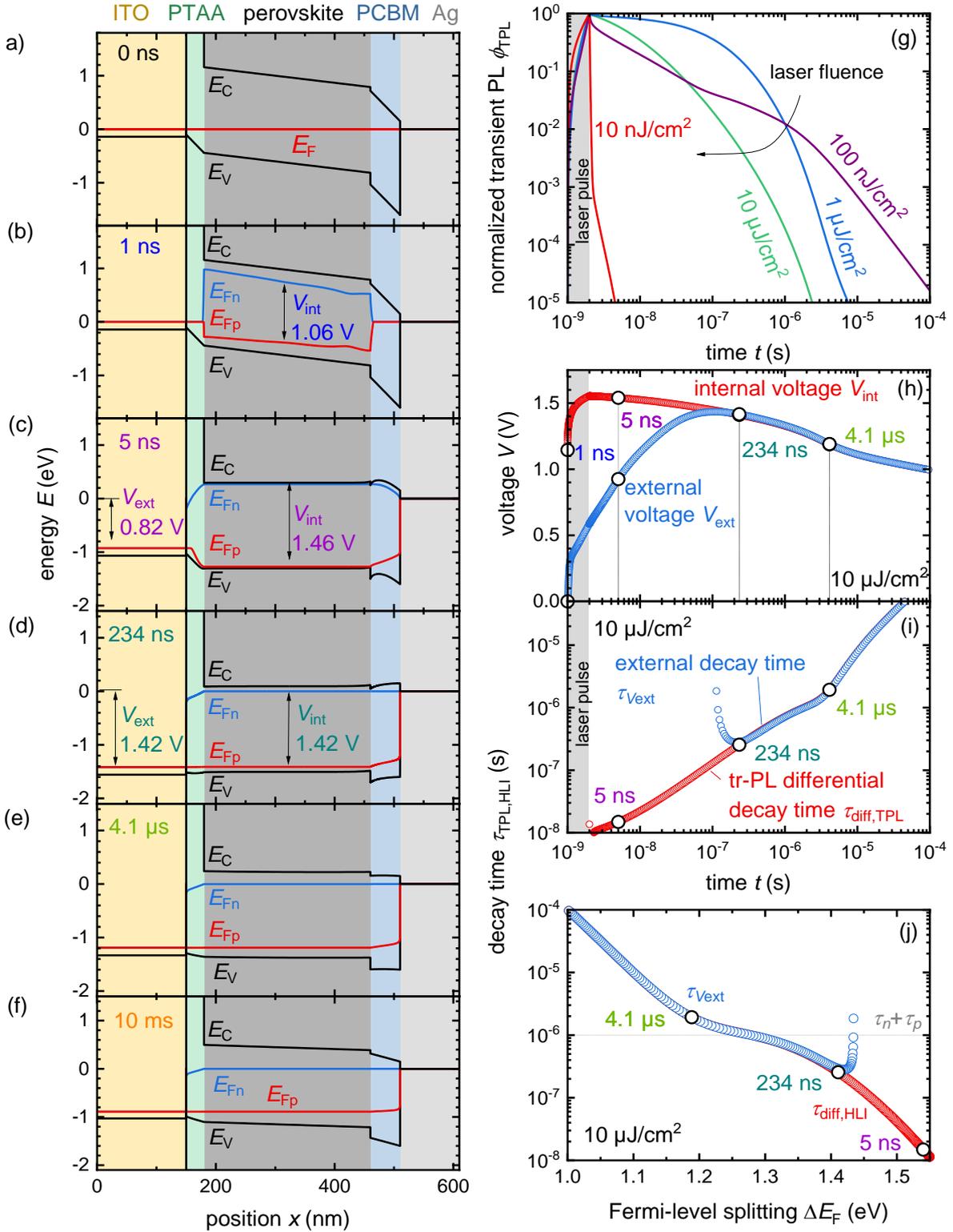

**Figure 9:** Numerical simulations results of a TPL experiment on a complete solar cell stack. (a-f) band diagrams of the glass/ITO/PTAA/perovskite/PCBM/Ag stack for different time delays. (g) normalized TPL transients for different laser fluences. (h) illustrates how the externally measurable voltage $V_{ext}$ between the contacts needs time to build up while the Fermi-level-splitting $\Delta E_{F,int} = qV_{int}$ inside the cell is already present. (i) shows the decay time $\tau_{TPL,HLI}$ as a function of time and (j) versus the Fermi-level splitting for the highest excitation fluence of 10 μJ/cm². As an illustration, we picked a cell that has good interface properties and a long bulk life. The entire simulation parameters can be found in Table S3 in the Supporting Information.

Figure 9d shows the situation, where the quasi-Fermi levels have equilibrated. Now the gradients $dE_{Fn}/dx$ and $dE_{Fp}/dx$ are nearly zero indicating that there is very little current flow to or from the electrodes. The internal and external voltage are equal, and both will decay over time in the same way. If we simulate the photoluminescence transients for the device geometry presented in Figure 9a-f, we observe a strong dependence of the decay curve on the pulse fluence of the laser. Figure 9g shows the decays normalized to their maximum. For the low pulse fluence of 10 nJ/cm$^2$, the PL decays quickly directly after the laser pulse has hit sample. In this case, nearly all electrons and holes generated by the laser pulse are necessary to change the amount of charge on the device electrodes for a sufficient external voltage to build up. If, however, the pulse energy is higher, no abrupt fast decay at early times is visible because only a small fraction of photogenerated electrons and holes are needed to build up the external voltage. However, the initial decay is still not monoexponential because Auger- and radiative recombination in the absorber will now matter at early times. The combination of these effects leads to a difficult to interpret situation, where fast decays at early times are visible for very low and very high pulse energies, however, for completely different reasons. Figure 9h summarizes the charging and discharging of the capacitance of the solar cell by showing how external and internal voltage change with time, which we explained before using the band diagrams. In Figure 9i-j we present the decay time versus time or Fermi-level splitting. For the complete solar cell device stack, the decay time $\tau_{\text{TPL,HLI}}$ vs. $\Delta E_F$ consists of three regions, which will be discussed in the following.

It is still possible to derive an analytical relation describing the transient if we assume high-level injection, include radiative recombination, SRH recombination and discharging of a constant capacitance. In this case, the solution of Equation 10 has to be expressed as time as a function of the Fermi-level splitting via

$$t = \left(1 - \frac{mk_BT}{q}\frac{C_{\text{area}}}{qd}k_{\text{rad}}\tau_{\text{SRH}}^{\text{eff}}\right)\tau_{\text{SRH}}^{\text{eff}} \ln\left[\frac{n(0) + n(t)n(0)k_{\text{rad}}\tau_{\text{SRH}}^{\text{eff}}}{n(t) + n(t)n(0)k_{\text{rad}}\tau_{\text{SRH}}^{\text{eff}}}\right] \\ + \frac{mk_BT}{q}\frac{C_{\text{area}}}{J_0}\left[\exp\left(-\frac{\Delta E_F(t)}{mk_BT}\right) - \exp\left(-\frac{\Delta E_F(0)}{mk_BT}\right)\right] \quad (11)$$

where $J_0 = qd_{\text{pero}}n_i/\tau_{\text{SRH}}^{\text{eff}}$, $n(t) = n_i \exp(\Delta E_F(t)/mk_BT)$ and $n(0)$ equals the initial charge-carrier density due to the generation of the laser pulse excitation. While we are not aware that Equation 11 has been previously presented in the literature, similar equations neglecting radiative recombination have been derived and used in the context of analysing OCVD decays, where the equation was used to derive a time dependent open-circuit voltage and subsequently a differential time constant of a large signal $V_{\text{oc}}$ decay.[40, 41, 43]

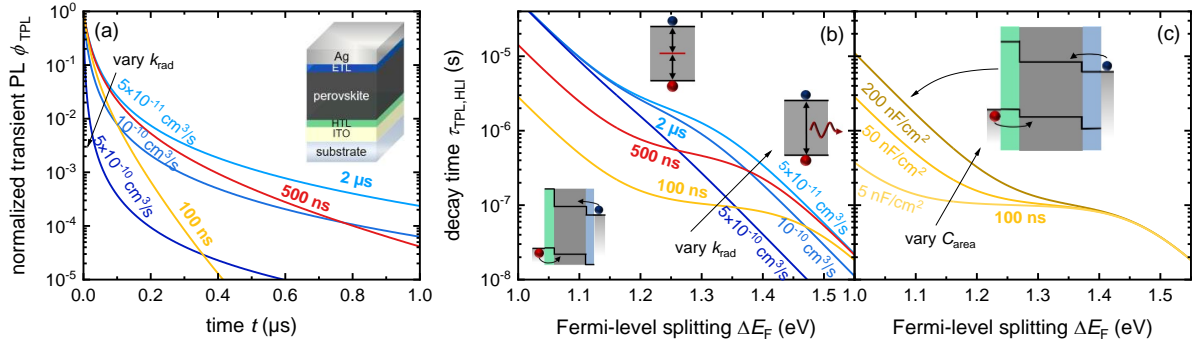

**Figure 10:** Illustration of (a) transient photoluminescence decays and (b-c) the decay time over Fermi-level splitting calculated from the analytical solution of the simplified differential equation in Equation 11 for the case of high-level injection and neglected Auger recombination, which can be used to demonstrate the charge-carrier dynamics in a perovskite solar cell. (a-b) Results for varying SRH bulk lifetimes $\tau_{SRH}^{bulk} = \tau_n + \tau_p$, namely 100 ns (yellow), 500 ns (red) and 2 µs (blue) are shown, as well as the effect due to changes of the radiative coefficients ($1 \times 10^{-10}$ cm$^3$s$^{-1}$, $5 \times 10^{-10}$ cm$^3$s$^{-1}$, $5 \times 10^{-11}$ cm$^3$s$^{-1}$). (c) variation and the capacitance (5 nF/cm$^2$, 50 nF/cm2, 200 nF/cm$^2$) and its impact on the shape of the decay time at low $\Delta E_F$. The discharging of the capacitances prolongs the decay time.

**Figure 10** illustrates Equation 11, whereby panel (a) shows the normalized TPL decays and panel (b) their decay times $\tau_{TPL,HLI}$ for three different SRH bulk lifetimes $\tau_{SRH}^{bulk} = \tau_n + \tau_p$, respectively 100 ns (yellow), 500 ns (red) and 2 µs (blue). Furthermore, the external radiative recombination coefficient $k_{rad}$ is varied for the SRH lifetime of 2 µs and the capacitance $C_{area}$ in calculations with $\tau_{SRH}^{bulk} = 100$ ns. With shorter SRH lifetimes and higher radiative coefficients, the PL transients show a faster decay, however, one can hardly note any systematic differences compared to the case of perovskite on glass (Figure 2a)). This picture changes when considering the decay times $\tau_{TPL,HLI}$ as a function of quasi-Fermi-level splitting in Figures 10b and c. In this case, three different regions appear in the decay time curve as was the case in Figure 9j. At early times and high Fermi-level-splitting, we observe a charge-carrier dependent lifetime that is indicative of radiative recombination, which is already familiar to us. At slightly longer times and lower Fermi-level-splitting $\tau_{TPL,HLI}$ reaches a plateau and becomes approximately constant at a value $\tau_{SRH}^{bulk} = \tau_n + \tau_p$. But instead of staying on or near this plateau as usual, the decay times for even smaller Fermi level splitting increase even further. This SRH dominated regime transitions into the exponentially voltage-dependent regime controlled by the ratio $C_{area}/J_0$. Here, charge-carrier recombination is limited by discharging of the electrodes. Thus, the decay time $\tau_{TPL,HLI}$ at small $\Delta E_F$ is higher for higher capacitances, demonstrated by the yellow curves. The purpose of transient photoluminescence measurements is to determine recombination parameters in finished devices. Thus, the region at low times and medium to high $\Delta E_F$ is the relevant region that one would like to assess and use for extraction of bulk lifetimes, surface recombination velocities and radiative recombination coefficients. This implies that the initial value $\Delta E_F(0)$ of the Fermi-level splitting and hence also the charge-carrier concentration $n(0)$ has to be high enough to observe the relevant range. Moreover, the different regimes can also merge into one another, so that a clear separation is no longer possible. This overlap occurs e.g. for a combination of high radiative recombination coefficients and high bulk lifetimes. Then the decay time $\tau_{\text{TPL,HLI}}$ is radiatively limited. The corresponding case is illustrated by the blue curves belonging to a SRH bulk lifetime of 2 µs. For $k_{\text{rad}} = 10^{-10}\,\text{cm}^3\text{s}^{-1}$ the plateau of $\tau_{\text{TPL,HLI}}$ is fully gone, and the decay time runs almost straight. At this point, it is no longer possible to determine whether there is a limitation due to capacitive effects or due to fast radiative recombination. This means that the determination of the recombination processes becomes particularly difficult for very good devices with long SRH lifetimes.

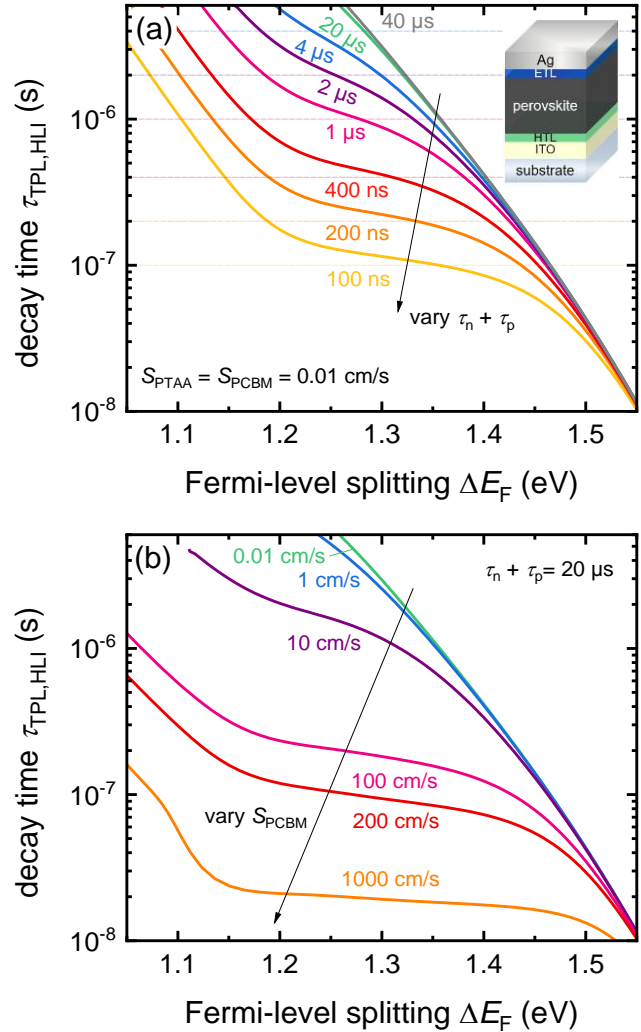

**Figure 11:** Decay times $\tau_{\text{TPL,HLI}}$ vs. Fermi-level splitting resulting from numerical simulations of a transient PL of solar cell stack using Sentaurus TCAD. (a) Simulation series where the SRH bulk lifetime is varied from 100 ns (yellow) to 40 µs (grey), while the interface recombination velocities are low ($S = 0.01$ cm/s) for both absorber-transport layer interfaces. In b) the opposite case is presented, where $\tau_n + \tau_p = 20\,\mu s$ is kept constant and the interface recombination velocity $S_{\text{PCBM}}$ of perovskite/PCBM is varied. Both changes translate into observable changes in the decay time. Additional simulation parameters are listed in Table S3 in the Supporting Information.

With **Figure 11** we confirm that the trend of the analytical relation can also be obtained in numerical simulations of a solar cell. In Figure 11a a variation of the SRH bulk lifetime (100 ns-40 μs) is illustrated. In this example, the interfaces between the perovskite absorber and the charge-extracting layers are nearly ideal (low $S$, no or small band offsets) and similar to the simulation parameters used in Figure 9 (see Table S3 in the Supporting Information). Here again, the SRH bulk lifetime correlates with the value of the decay time $\tau_{TPL,HLI}$ near the inflection point of the curve and also the radiative limitation for high $\tau_n + \tau_p$ becomes apparent. A direct comparison of the analytical and numerical solution can be found in Figure S14 in the Supporting Information. Figure 11b deals with the case of varying interface-recombination velocities $S_{PCBM}$ on the perovskite/PCBM side, while the SRH bulk lifetime is high, namely $\tau_n + \tau_p = 20$ μs. For higher $S_{PCBM}$ the decay time in the middle region around 1.2 to 1.4 eV drops.

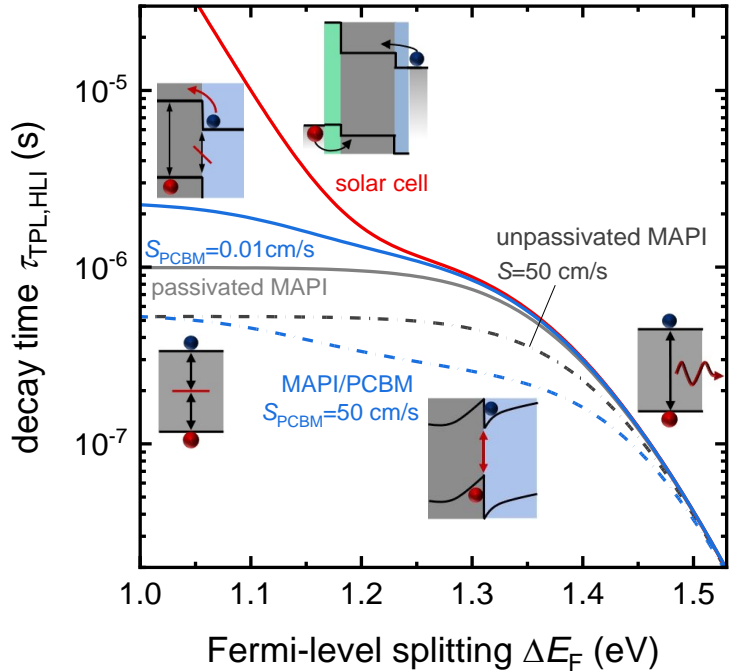

**Figure 12:** Summary of the fundamentally different effects that modify the decay time in different sample geometries. Bulk recombination (radiative and SRH) is the sole factor influencing the decay in a passivated perovskite layer (solid gray line)– the simplest possible sample geometry discussed here. In the presence of surface recombination (dot-dashed dark grey line) the decay time $\tau_{TPL,HLI}$ at low $\Delta E_F$ is reduced. In case of a perovskite/ETL bilayer (blue curves), the shape of the decay time is fundamentally changed due to effects such as interface charging (dash-dotted blue line) and electron reinjection (solid blue line) from the ETL into the perovskite. In addition to interfacial charging, we also observe electrode charging and discharging if we analyze complete devices (red line). In this case, we observe an exponential increase of $\tau_{TPL,HLI}$ towards lower values of $\Delta E_F$.

Finally, **Figure 12** shows a summary of some exemplary effects observed so far in the simulations and how they affect the charge-density-dependent decay time. The simplest case of a passivated perovskite film on glass (solid grey curve) shows fast radiative recombination at high $\Delta E_F$ and the bulk lifetime of SRH recombination at low $\Delta E_F$. For less well passivated surfaces, the $\tau_{TPL,HLI}$-curve is essentially only shifted to lower decay times at lower values of $\Delta E_F$ (dot-dashed grey curve). An additional dip is observed for absorber-charge transfer layer combinations (dot-dashed blue curve). Here, interface charging and Coulomb induced charge accumulation at the MAPI/PCBM interface lead to increased interface recombination and decay times $\tau_{TPL,HLI}$ that are reduced relative to the case of a MAPI layer with the same surface recombination velocity but without the PCBM layer attached. For perovskite/ETL bilayers with substantial conduction-band offset but recombination inactive interface, we would observe additional effects due to electron re-injection into the perovskite. This re-injection depends on the energetic barrier at the perovskite/ETL interface and leads to $\tau_{TPL,HLI}$ that can substantially exceed the bulk SRH lifetimes of the perovskite itself (solid blue line). Finally, we can also charge and discharge the electrode capacitance of a complete perovskite solar cell (solid red line). At open circuit, the capacitor can only discharge via recombination in the absorber layer or at its interfaces. Since the recombination resistance of the solar cell increases exponentially towards lower voltages, the decay time associated with electrode discharging also increases exponentially.

## 6. Comparison to experimental data

Finally, we want to apply and transfer our findings from the simulations to experimental data. **Figure 13** shows experimental data obtained from photoluminescence measurements for two sample series each featuring different multilayer stacks from the film on glass to the complete device. One sample series is based on solution-processed MAPI (left side), while the second sample series uses coevaporated MAPI ($CH_3NH_3PbI_3$) as absorber material (right side). The used stacks for the full devices are glass/ITO/PTAA/MAPI/ETL/BCP/Ag, where ETL is PCBM for the cell with solution-processed MAPI and $C_{60}$ for the cell with coevaporated MAPI. Note that the open-circuit voltages of the two solar cells are substantially different despite the fact that the stack and the materials are nearly identical. The coevaporated, inverted, planar MAPI solar cell has an open-circuit voltage $V_{oc}$ of only 1.05 V while the solution-processed MAPI solar cell has a very high open-circuit voltage of 1.25 V (see Supporting Figure S18). The sample series, we show, always include a sample (glass/MAPI/TOPO) that should be well passivated. The molecule n-trioctylphosphine oxide (TOPO) has been shown to strongly reduce surface recombination velocities.[26, 27] In addition, the sample series include one sample with the hole transport layer PTAA (glass/PTAA/MAPI/TOPO) that serves to characterize recombination losses at the PTAA/MAPI interface. Finally, the complete cells are included which feature in addition the MAPI/ETL interface as an additional source of recombination. Figure 13a and b compare the normalized transient PL intensities $\phi_{TPL}$ and (c-d) the respective decay times $\tau_{TPL,HLI}$ vs. $\Delta E_F$ for the two sample series described above. Each data set results from stitching several measurements recorded with a gated CCD camera starting at different delay times after the laser pulse and using different gains and integration times. This approach enables a very high dynamic range of up to 7 orders of magnitude that is necessary to observe the wide range of physical phenomena discussed in the simulation sections of this paper. Typical combined measurement times were several hours. Given the differences in $V_{oc}$ of the two cells, we expect that the two sample series should differ in the recombination losses that occur either in the bulk or at interfaces and that these differences should be reflected in differences in the decay times $\tau_{TPL,HLI}$. This qualitative expectation is confirmed already by studying the PL transients for the two sample series (Figure 13a and b) which show substantially faster TPL decays for the samples based on coevaporated perovskite layers.

In order to obtain additional insights, we first determine the decay times shown in Figure 13c and d. While taking the derivative defined by Equation 3 is a simple task for smooth, simulated data, it is quite challenging to extract meaningful derivatives from noisy experimental data. Therefore, it is advisable to not or not only take the derivative of the background corrected raw data but to first fit the data with a function and then differentiate the fit. We note that given the multitude of non-exponential features affecting the transients, we opted to identify functions for which the fit algorithm converges easily and leads to a good agreement with the experimental transients even though the functional form of the fit functions bears no physical meaning. We observed that good candidates for fit functions are high order polynomials that can be fitted to the logarithm of the PL. Alternatives are rational functions, i.e. ratios of higher order polynomials. In Figure 13c and d, we therefore see symbols and lines, where the symbols represent the derivative of the background-corrected and stitched raw data while the lines are the derivative of the fits to the data. Both agree within the accuracy of the method, which is – at least partly – a consequence of the raw data being fairly high quality (low noise level) due to long integration and high measurement times. The $\Delta E_F$ axis for the experimental data was determined from the knowledge of the laser fluences used in the measurement (provides the initial $\Delta E_F$ at time zero after the pulse with Equation 4) and the knowledge that the PL intensity scales with $\exp(\Delta E_F/k_BT)$.[47] This proportionality implies that any order of magnitude decrease in PL leads to a 58 mV decrease in $\Delta E_F$. From comparing the glass/MAPI/TOPO (grey) samples of the coevaporated and the solution-processed perovskite sample series in Figure 13c and d, it is directly apparent that the quality of the bulk differs substantially. In addition, the coevaporated cell suffers from increased interface recombination relative to the solution-processed cell, which we conclude from the substantial reduction in $\tau_{TPL,HLI}$ for the samples with interfaces (green and red) relative to the passivated layer on glass (grey). Furthermore, the comparison in Figure 13 demonstrates that the representation of the decay time $\tau_{TPL,HLI}$ via Fermi-level-splitting $\Delta E_F$ is advantageous compared to the usual representation of the decay itself. This new type of graph highlights differences and similarities between the samples more clearly and allows estimating recombination parameters.

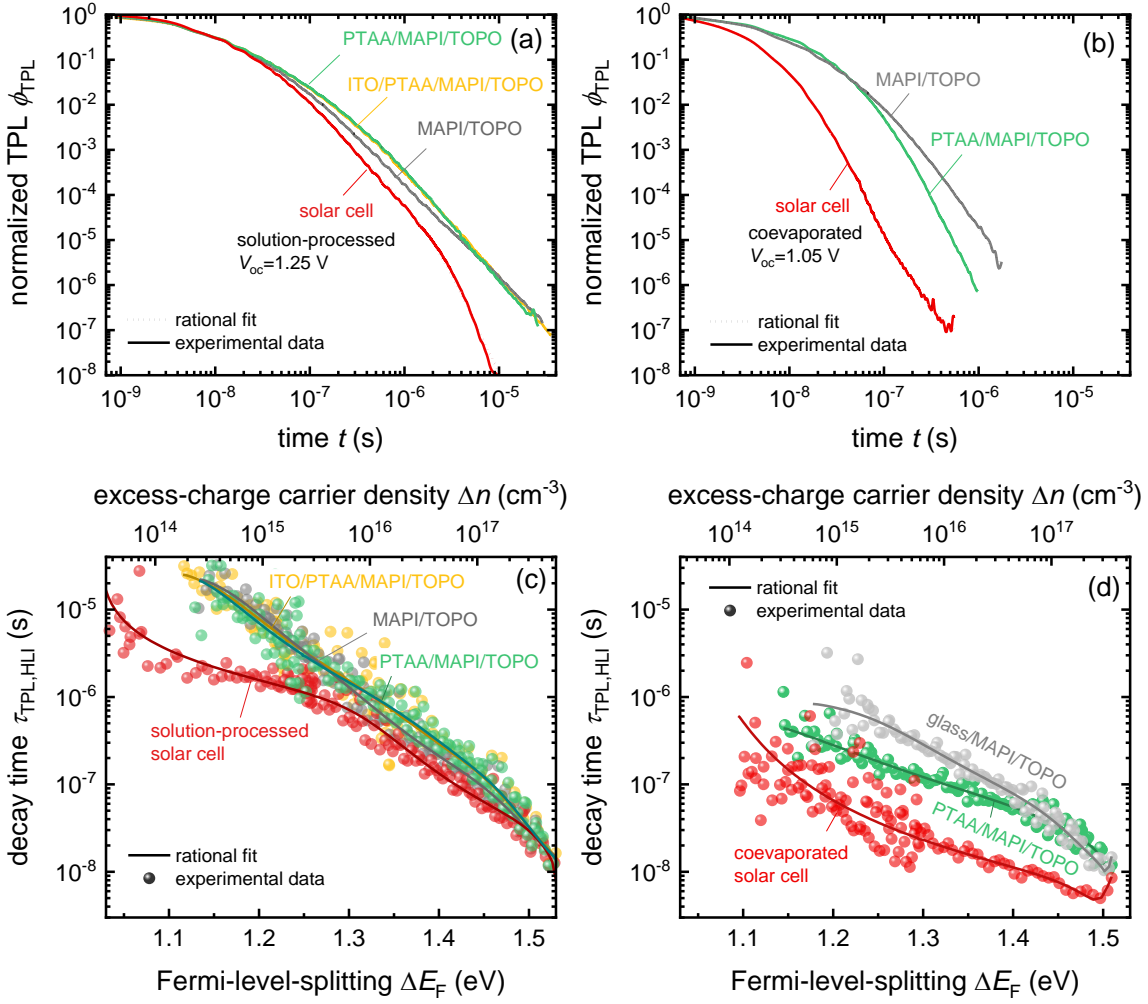

**Figure 13:** Experimental data of transient photoluminescence measurements on two different a sample series comparing coevaporated and solution-processed, starting with the passivated perovskite film (grey), going further to MAPI with charge-extracting layer (PTAA (green), $C_{60}$ (blue)) and finally showing the complete coevaporated cell (red). A laser excitation fluences of 7.2 μJ/cm² with a spot diameter of 3.8 mm was used for the measurements. A gated CCD camera is used to detect the time-resolved PL. Stitching several measurements recorded at different delay times after the laser pulse with different gains and integration times enables a higher dynamic range. More details can be found in the method section in the Supporting Information. (a-b) Normalized photoluminescence decays over time. (c-d) Decay time $\tau_{TPL,HLI}$ assuming HLI ($m=2$) versus Fermi-level splitting $\Delta E_F$. Solid lines represent the derivative from TPL fits using a fifth-degree rational function and are intended to be as a guide to the eye.

To quantify and extract these material and device parameters, the TPL datasets were fitted with transient simulations performed with Sentaurus TCAD. **Figure 14** presents the experimental data for $\tau_{TPL,HLI}$ of the two sample series compared with the simulations that best reproduce the experimental data. First, we simulated the glass/perovskite/TOPO stacks (grey) to determine the recombination coefficients of the perovskite bulk material. The decay time $\tau_{TPL,HLI}$ of experimental data is dominated by radiative recombination over the complete range of quasi Fermi-level splitting, which were experimentally accessible, and increases constantly for smaller $\Delta E_F$. This behaviour implies that SRH lifetimes have to be extremely long. The simulations are done with a SRH bulk lifetime of $\tau_n + \tau_p = 80$ µs and in Figure S21, we show how the simulations with lower SRH lifetimes look like. From Figure S21, we conclude for lifetimes $\tau_n + \tau_p < 40$ µs, the agreement between simulation and experiment deteriorates substantially. SRH bulk lifetimes as high as 80 µs would allow an open-circuit voltage of 1.31 V under AM1.5g illumination, i.e. a value very close to the radiative limit of 1.32 V.[13] The solution-processed samples with PTAA (green, yellow) do not suffer from additional losses. The decay times are similar to the passivated sample with the glass/perovskite interface. The band offset between perovskite and PTAA and the surface recombination velocity must therefore be negligibly small. In the Supporting Figure S21 we demonstrate that a surface recombination velocity must be around $S_{PTAA}=1$ cm/s. Also, the band offset and surface recombination at the PCBM/perovskite interface must be quite small to explain the data. Nevertheless, the PCBM/perovskite interface is the only interface that causes visible deviations of the PL transients from the behaviour expected in the radiative limit. We find that a surface recombination velocity $S_{PCBM}=17$ cm/s and an offset of 70 meV lead to the best agreement with the experimental data. Note that the decay time curve for the solution-processed cell (red line, Figure 12c) reproduces the typical S-shape predicted by Equation 11. It should also be noted that the simulated $JV$-curve of a simulated cell with the stated parameters agrees well with the measured $JV$-characteristic (Supporting Figure S19). Another implication from the data shown in Figure 14a is that for perovskite films and layer stacks with small recombination losses, the measured differential time constants may assume nearly any value (from tens of ns to tens of ms), depending on the range of carrier concentrations and Fermi-level splitting that are set by the laser fluence.

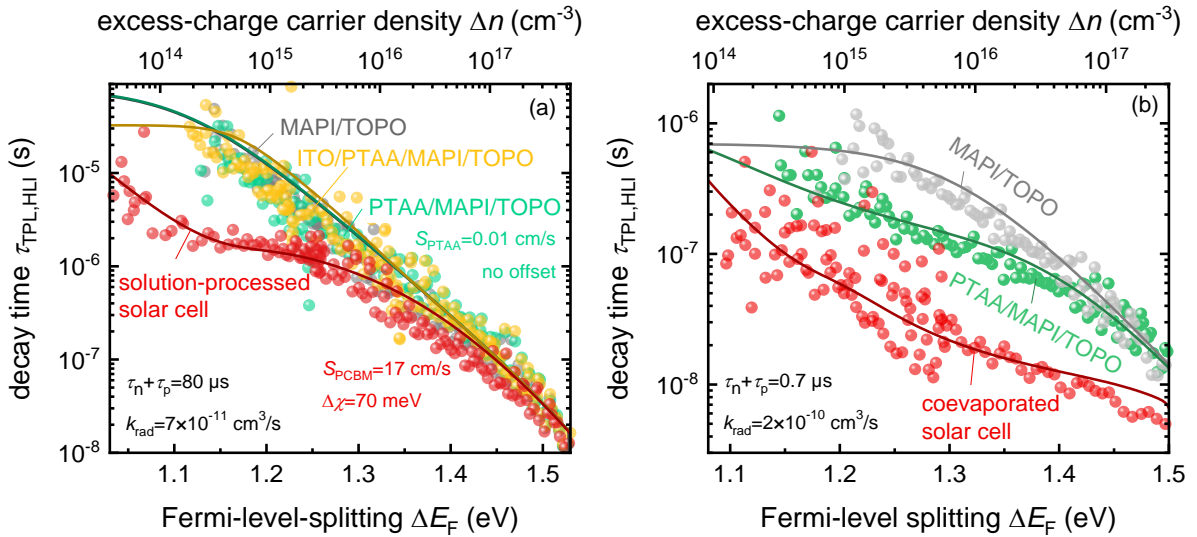

**Figure 14:** Experimental data of transient photoluminescence measurements of the solution-processed sample series and fits from Sentaurus TCAD, which allow us to state the material parameters that describe the sample behavior best. All simulation parameters are listed in Table S4 and S5 in the Supporting Information.

Without the additional information on the laser fluence and without high dynamic range data as shown here, the information obtained from TPL data on many high-quality layers or layer stacks would be either difficult to compare or entirely meaningless.

In the following we want to discuss the fitting of the coevaporated sample series, representing an example of samples with a higher degree of non-radiative recombination. A SRH bulk lifetime of about $\tau_n + \tau_p = 750$ ns best describes the experimental TPL data of coevaporated bulk passivated with TOPO in the glass/perovskite/TOPO stack (grey). This SRH bulk lifetime would allow much higher open-circuit voltages than 1.05 V. Combined with an effective radiative recombination coefficient $k_{rad} = 2\times10^{-10}$ cm$^3$s$^{-1}$, a $V_{oc}$ of about 1.23 V would still be possible for the coevaporated bulk if no additional recombination losses would occur in the stack. The simulations suggest that for the coevaporated cell, these additional recombination losses are caused by misaligned energy levels and increased recombination at both interfaces (see Figure 6). Interface recombination at the PTAA/MAPI interface (coevaporated sample) leads to slightly shorter decay times $\tau_{TPL,HLI}$ at small Fermi-level splitting (green) as opposed to the samples without charge-extracting layers attached (grey). Note that these two decay time curves nicely overlap at high $\Delta E_F$, where radiative and Auger recombination dominate the $\tau_{TPL,HLI}$. The trend of PTAA/perovskite interface can be best explained by a band offset of $\Delta\chi = 100$ meV and a surface recombination velocity of $S_n = S_p = 100$ cm/s. Since the PTAA layer is very thin we would expect that $\tau_{TPL,HLI}$ saturates for small $\Delta E_F$ around ~400 ns (discussion section 3). However, the decay time at small $\Delta E_F$ is much higher and increases beyond the bulk lifetime. We observe this behaviour in our simulations if we assume shallow, neutral defects. Thus, in the simulation that best fits the data the defects are shallow and positioned only 150 meV away from the VB band edge (Supporting Figure S22). These parameters would still allow a Fermi-level splitting of 1.18 V in simulated PTAA/perovskite stack. Also, the properties of the perovskite/C$_{60}$ interface in the solar cell (red) are quite complex. In this case, the decay time is roughly two orders of magnitudes lower suggesting the additional interfacial losses in $V_{oc}$ must be higher. Only for small $\Delta E_F$ the decay time increases again indicating that the loss is caused by a huge conduction band offset e.g. $\Delta\chi = 200$ meV. To explain the shape of $\tau_{TPL,HLI}$ vs. $\Delta E_F$ that deviates from the typical S-shape we introduced in the discussion of the TPL on solar cells, we had to assume at least two different defect state. A deep level defect is responsible for the loss in $V_{oc}$ and another shallow defect must be introduced to match the shape of the differential decay curve. While we note here that various defect properties such as its charge, concentration and energetic and spatial position affect the TPL, it is beyond the scope of this work to present a systematic investigation on the influence of these parameters on the decay time $\tau_{TPL,HLI}$.

## 7. Conclusions

Transient photoluminescence experiments are abundantly used in the field of halide perovskite photovoltaics to study charge-carrier recombination in the bulk and at interfaces. While the interpretation of TPL on thin films on glass has been thoroughly discussed and used in the literature,[12, 13, 48] the most important recombination losses are often occurring at the interfaces between the absorber and the charge-transfer layers.[15] In addition, the presence of charge-transfer layers can also have an impact on how the perovskite films grow and hence affect the bulk and surface quality of the perovskite layer. Thus, there is a clear need to extend our theoretical understanding to TPL measurements done on a variety of sample geometries including zero, one or two charge-transfer layers in contact to the perovskite absorber. In addition, it is important to also understand how contact layers such as ITO or Ag affect the TPL decay and to be able to understand measurements done on complete devices. The present paper provides an extensive account based on a combination of experimental data, numerical simulations with Sentaurus TCAD and analytical solutions to differential equations that allows the reader to understand the mechanisms affecting a TPL decay in a variety of sample geometries. As a key tool to analyze the data, we introduce the concept of a decay time $\tau_{TPL,HLI}$ displayed as a function of the time-dependent quasi-Fermi level splitting. This plot allows us to combine data obtained with different laser fluences in one figure and improves the comparability of different data sets. The charge-carrier-density dependent decay time $\tau_{TPL,HLI}$ is affected by various different *recombination mechanisms* (radiative, Auger and SRH recombination), also by charge transfer between the absorber and the charge-transport layers and, finally, by capacitive charging and discharging of the electrodes in case of a full device. The different mechanisms can partly be distinguished by their appearance at different values of the quasi-Fermi level splitting and by their characteristic slope the decay time vs. quasi-Fermi level splitting diagram. Along with this paper, the reader finds a video collection of the band diagram during the TPL simulation for the different layer stacks. After introducing the general concepts using numerical simulations, we show experimental data sets on different sample geometries and absorber deposition methods (solution-processed vs. coevaporated). We determine the TPL decays over seven orders of magnitude in dynamic range and show that our previously presented recipe for MAPI layers allows bulk lifetimes of several tens of μs (best fits are obtained for 80 μs the sum of electron and hole lifetimes). In addition, the TPL transients clearly indicate negligible losses at the MAPI/PTAA interface and only moderate losses with surface recombination velocities of 17 cm/s for the MAPI/PCBM interface.

**Data availability**

The data that support the findings of this study are available from the corresponding author on request.

**Code availability**

The code and data sets generated and analyzed during this study are available from the corresponding author on reasonable request.

**Author Information**

Corresponding Author:

*To whom correspondence should be addressed. E-Mail: t.kirchartz@fz-juelich.de, l.krueckemeier@fz-juelich.de

**Acknowledgements**

The authors acknowledge support from PEROSEED. We thank Steffen Krause and Jonathan Werner for the process development and fabrication of the coevaporated MAPI samples.

**Author contributions**

LK and TK prepared the manuscript. LK performed the simulations. BK developed the basis of the Sentaurus TCAD code, which was modified by LK. ZL fabricated the solar cell devices and thin films. LK performed the TPL measurements. All authors contributed to the planning and the interpretation of simulations and experiments. All authors discussed the results and commented on the manuscript.

**References**

[1] W. E. I. Sha, H. Zhang, Z. S. Wang, H. L. Zhu, X. Ren, F. Lin, A. K.-Y. Jen, W. C. H. Choy, *Advanced Energy Materials* **2018,** 8, 1701586.
[2] S. Draguta, J. A. Christians, Y. V. Morozov, A. Mucunzi, J. S. Manser, P. V. Kamat, J. M. Luther, M. Kuno, *Energy & Environmental Science* **2018,** 11, 960.
[3] L. Krückemeier, U. Rau, M. Stolterfoht, T. Kirchartz, *Advanced Energy Materials* **2020,** 10, 1902573.
[4] P. K. Nayak, S. Mahesh, H. J. Snaith, D. Cahen, *Nature Reviews Materials* **2019,** 4, 269.
[5] M. A. Green, A. W. Ho-Baillie, *ACS Energy Letters* **2019,** 4, 1639.
[6] J. Yao, T. Kirchartz, M. S. Vezie, M. A. Faist, W. Gong, Z. He, H. Wu, J. Troughton, T. Watson, D. Bryant, J. Nelson, *Physical Review Applied* **2015,** 4, 014020.
[7] K. Tvingstedt, O. Malinkiewicz, A. Baumann, C. Deibel, H. J. Snaith, V. Dyakonov, H. J. Bolink, *Scientific Reports* **2014,** 4, 6071.
[8] G.-J. A. H. Wetzelaer, M. Scheepers, A. M. Sempere, C. Momblona, J. Ávila, H. J. Bolink, *Advanced Materials* **2015,** 27, 1837.
[9] D. Luo, R. Su, W. Zhang, Q. Gong, R. Zhu, *Nature Reviews Materials* **2020,** 5, 44.
[10] T. Kirchartz, J. A. Márquez, M. Stolterfoht, T. Unold, *Advanced Energy Materials* **2020,** 10, 1904134.
[11] K. P. Goetz, A. D. Taylor, F. Paulus, Y. Vaynzof, *Advanced Functional Materials* **2020,** 30, 1910004.
[12] S. D. Stranks, V. M. Burlakov, T. Leijtens, J. M. Ball, A. Goriely, H. J. Snaith, *Physical Review Applied* **2014,** 2, 034007.
[13] F. Staub, H. Hempel, J. C. Hebig, J. Mock, U. W. Paetzold, U. Rau, T. Unold, T. Kirchartz, *Physical Review Applied* **2016,** 6, 044017.
[14] J. M. Richter, M. Abdi-Jalebi, A. Sadhanala, M. Tabachnyk, J. P. H. Rivett, L. M. Pazos-Outon, K. C. Gödel, M. Price, F. Deschler, R. H. Friend, *Nature Communications* **2016,** 7, 13941.
[15] M. Stolterfoht, P. Caprioglio, C. M. Wolff, J. A. Marquez, J. Nordmann, S. Zhang, D. Rothhardt, U. Hörmann, Y. Amir, A. Redinger, L. Kegelmann, F. Zu, S. Albrecht, N. Koch, T. Kirchartz, M. Saliba, T. Unold, D. Neher, *Energy & Environmental Science* **2019,** 12, 2778.
[16] M. Stolterfoht, C. M. Wolff, J. A. Marquez, S. Zhang, C. J. Hages, D. Rothhardt, S. Albrecht, P. L. Burn, P. Meredith, T. Unold, D. Neher, *Nature Energy* **2018,** 3, 847.
[17] P. Schulz, D. Cahen, A. Kahn, *Chemical Reviews* **2019,** 119, 3349.
[18] C. M. Wolff, P. Caprioglio, M. Stolterfoht, D. Neher, *Advanced Materials* **2019,** 31, 1902762.
[19] J. Chen, N.-G. Park, *Advanced Materials* **2019,** 31, 1803019.
[20] M. Stolterfoht, M. Grischek, P. Caprioglio, C. M. Wolff, E. Gutierrez-Partida, F. Peña-Camargo, D. Rothhardt, S. Zhang, M. Raoufi, J. Wolansky, M. Abdi-Jalebi, S. D. Stranks, S. Albrecht, T. Kirchartz, D. Neher, *Advanced Materials* **2020,** 32, 2000080.
[21] J. Haddad, B. Krogmeier, B. Klingebiel, L. Krückemeier, S. Melhem, Z. Liu, J. Hüpkes, S. Mathur, T. Kirchartz, *Advanced Materials Interfaces* **2020,** 7, 2000366.
[22] B. Krogmeier, F. Staub, D. Grabowski, U. Rau, T. Kirchartz, *Sustainable Energy & Fuels* **2018,** 2, 1027.
[23] E. M. Hutter, T. Kirchartz, B. Ehrler, D. Cahen, E. v. Hauff, *Applied Physics Letters* **2020,** 116, 100501.
[24] V. Sarritzu, N. Sestu, D. Marongiu, X. Chang, S. Masi, A. Rizzo, S. Colella, F. Quochi, M. Saba, A. Mura, *Scientific reports* **2017,** 7, 44629.
[25] M. Stolterfoht, V. M. Le Corre, M. Feuerstein, P. Caprioglio, L. J. A. Koster, D. Neher, *ACS Energy Letters* **2019,** 4, 2887.
[26] D. W. DeQuilettes, S. Koch, S. Burke, R. K. Paranji, A. J. Shropshire, M. E. Ziffer, D. S. Ginger, *ACS Energy Letters* **2016,** 1, 438.
[27] I. L. Braly, D. W. deQuilettes, L. M. Pazos-Outón, S. Burke, M. E. Ziffer, D. S. Ginger, H. W. Hillhouse, *Nature Photonics* **2018,** 12, 355.
[28] Y. Yang, M. Yang, D. T. Moore, Y. Yan, E. M. Miller, K. Zhu, M. C. Beard, *Nature Energy* **2017,** 2, 1.
[29] L. M. Herz, *ACS Energy Letters* **2017,** 2, 1539.
[30] X. Zhang, J.-X. Shen, W. Wang, C. G. Van de Walle, *ACS Energy Letters* **2018,** 3, 2329.

[31] C. L. Davies, M. R. Filip, J. B. Patel, T. W. Crothers, C. Verdi, A. D. Wright, R. L. Milot, F. Giustino, M. B. Johnston, L. M. Herz, *Nature Communications* **2018,** 9, 293.
[32] J.-X. Shen, X. Zhang, S. Das, E. Kioupakis, C. G. Van de Walle, *Advanced Energy Materials* **2018,** 8, 1801027.
[33] B. Das, I. Aguilera, U. Rau, T. Kirchartz, *Physical Review Materials* **2020,** 4, 024602.
[34] X. Zhang, M. E. Turiansky, J.-X. Shen, C. G. Van de Walle, *Physical Review B* **2020,** 101, 140101.
[35] T. Kirchartz, F. Staub, U. Rau, *ACS Energy Letters* **2016,** 1, 731.
[36] T. Kirchartz, *Philosophical Transactions of the Royal Society A: Mathematical, Physical and Engineering Sciences* **2019,** 377, 20180286.
[37] C. G. Shuttle, B. O'Regan, A. M. Ballantyne, J. Nelson, D. D. C. Bradley, J. d. Mello, J. R. Durrant, *Applied Physics Letters* **2008,** 92, 093311.
[38] S. Wheeler, D. Bryant, J. Troughton, T. Kirchartz, T. Watson, J. Nelson, J. R. Durrant, *The Journal of Physical Chemistry C* **2017,** 121, 13496.
[39] P. R. F. Barnes, K. Miettunen, X. Li, A. Y. Anderson, T. Bessho, M. Gratzel, B. C. O'Regan, *Advanced Materials* **2013,** 25, 1881.
[40] L. Castaner, E. Vilamajo, J. Llaberia, J. Garrido, *Journal of Physics D: Applied Physics* **1981,** 14, 1867.
[41] D. Kiermasch, A. Baumann, M. Fischer, V. Dyakonov, K. Tvingstedt, *Energy & Environmental Science* **2018,** 11, 629.
[42] F. A. Lindholm, J. J. Liou, A. Neugroschel, T. W. Jung, *IEEE Transactions on Electron Devices* **1987,** 34, 277.
[43] O. J. Sandberg, K. Tvingstedt, P. Meredith, A. Armin, *The Journal of Physical Chemistry C* **2019,** 123, 14261.
[44] Z. S. Wang, F. Ebadi, B. Carlsen, W. C. H. Choy, W. Tress, *Small Methods* **2020,** 4, 2000290.
[45] F. Steiner, S. Foster, A. Losquin, J. Labram, T. D. Anthopoulos, J. M. Frost, J. Nelson, *Materials Horizons* **2015,** 2, 113.
[46] Z. Liu, L. Krückemeier, B. Krogmeier, B. Klingebiel, J. A. Márquez, S. Levcenko, S. Öz, S. Mathur, U. Rau, T. Unold, T. Kirchartz, *ACS Energy Letters* **2019,** 4, 110.
[47] P. Würfel, *Journal of Physics C: Solid State Physics* **1982,** 15, 3967.
[48] M. J. Trimpl, A. D. Wright, K. Schutt, L. R. V. Buizza, Z. Wang, M. B. Johnston, H. J. Snaith, P. Müller-Buschbaum, L. M. Herz, *Advanced Functional Materials* **2020,** 30, 2004312.